\documentclass[universe,article,accept,pdftex,moreauthors]{Definitions/mdpi}

\firstpage{1}
\pubvolume{0}
\issuenum{0}
\articlenumber{0}
\pubyear{2024}
\copyrightyear{2024}
\externaleditor{Academic Editor: Name}
\datereceived{26 December 2023}
\daterevised{2 February 2024} % Comment out if no revised date
\dateaccepted{12 March 2024}
\datepublished{ }
\hreflink{https://doi.org/} % If needed use \linebreak
\Title{Constraining the Initial Mass Function in the Epoch of Reionization from Astrophysical and Cosmological Data}

\TitleCitation{Constraining the Initial Mass Function in the Epoch of Reionization from Astrophysical and Cosmological Data}

\Author{Andrea Lapi $^{1,2,3,4,}$*\orcidA{}, Giovanni Gandolfi $^{1,2,3}$\orcidB{}, Lumen Boco $^{1,2,3}$\orcidC{}, Francesco Gabrielli $^{1}$\orcidD{}, Marcella Massardi $^{1,4}$\orcidE{}, Balakrishna S. Haridasu $^{1,2,3}$\orcidF{}, Carlo Baccigalupi $^{1,2,3,5}$\orcidG{}, Alessandro Bressan $^{1,2}$\orcidH{} and Luigi Danese $^{1,2}$\orcidI{}}
\AuthorCitation{Lapi, A.; Gandolfi, G.; Boco, L.; Gabrielli, F.; Massardi, M.; Haridasu, B.S.; Baccigalupi, C.; Bressan, A.; Danese, L.}

\address{$^{1}$ \quad Scuola Internazionale Superiore Studi Avanzati (SISSA), Physics Area, Via Bonomea 265, 34136 Trieste, Italy; ggandolfi@sissa.it (G.G.); lboco@sissa.it (L.B.); 	fgabriel@sissa.it (F.G.); massardi@ira.inaf.it (M.M.); sandeep.haridasu@sissa.it (B.S.H.); bacci@sissa.it (C.B.); sbressan@sissa.it (A.B.); danese@sissa.it (L.D.)\\
$^{2}$ \quad IFPU-Institute for Fundamental Physics of the Universe, Via Beirut 2, 34014 Trieste, Italy\\ $^{3}$ \quad INFN-Sezione di Trieste, Via Valerio 2, 34127 Trieste,  Italy\\ $^{4}$ \quad IRA-INAF, Via Gobetti 101, 40129 Bologna, Italy\\
$^{5}$ \quad INAF, Osservatorio Astronomico di Trieste, Via G. B. Tiepolo 11, 34131 Trieste, Italy}

\corres{Correspondence: lapi@sissa.it}

\abstract{We aim to constrain the stellar initial mass function (IMF) during the epoch of reionization. To this purpose, we build up a semi-empirical model for the reionization history of the Universe based on various ingredients: the latest determination of the UV galaxy luminosity function from JWST out to redshift $z\lesssim 12$; {data-inferred and simulation-driven} assumptions on the redshift-dependent escape fraction of ionizing photons from primordial galaxies; a simple yet flexible parameterization of the IMF $\phi(m_\star)\sim m_\star^\xi\, e^{-m_{\star,\rm c}/m_\star}$ in terms of a high-mass end slope $\xi<0$ and a characteristic mass $m_{\star,\rm c}$, below which a flattening or a bending sets in {(allowing description of a variety of IMF shapes from the classic Salpeter to top-heavy ones)}; the \texttt{PARSEC} stellar evolution code to compute the UV and ionizing emission from different stars' masses as a function of age and metallicity; and a few physical constraints related to stellar and galaxy formation in faint galaxies at the reionization redshifts. We then compare our model outcomes with the reionization observables from different astrophysical and cosmological probes and perform Bayesian inference on the IMF parameters via a standard MCMC technique. We find that the IMF slope $\xi$ is within the range from $-2.8$ to $-2.3$, consistent with direct determination from star counts in the Milky Way, while appreciably flatter slopes are excluded at great significance. However, the bestfit value of the IMF characteristic mass $m_{\star,\rm c}\sim$ a few $M_\odot$ implies a suppression in the formation of small stellar masses at variance with the IMF in the local Universe. {This may be induced by} the thermal background of $\sim$20--30 K provided by CMB photons at the reionization redshifts. We check that our results are robust against different parameterizations for the redshift evolution of the escape fraction. Finally, we investigate the implications of our reconstructed IMF for the recent JWST detections of massive galaxies at and beyond the reionization epoch, showing that any putative tension with the standard cosmological framework is substantially alleviated.}


\begin{document}

\section{Introduction}\label{sec|intro}

One of the very basic ingredients for understanding galaxy formation and evolution is constituted by the stellar initial mass function or IMF $\phi(m_\star)$, which describes the distribution of stars' masses $m_\star$ per unit solar mass formed. Unfortunately, the IMF is largely uncertain, both from an observational and from a theoretical point of view (see \cite{Bastian10,Krumholz14,Kroupa21} and references therein).

Based on direct star counts in the Milky Way and its satellite galaxies, the IMF shape has been historically characterized in terms of a power law $\phi(m_\star)\propto m_\star^{-2.3}$ at the high-mass end \cite{Salpeter55}, going below $m_\star\approx 1\, M_\odot$ into a flatter power law $\phi(m_\star)\propto m_\star^{-1.3}$ \cite{Kroupa01} or into a log-normal shape \cite{Chabrier03}. An exponential downturn has been also claimed to occur, especially for metal-poor stars \cite{Larson98,Wise12}. Moreover, there is an ongoing debate in the literature on whether the IMF shape depends on the properties of the galactic environments where star formation proceeds. Redshift, age, metallicity, density/temperature of the interstellar medium, (specific) star formation rate (both volume and surface density), galaxy type, etc., have been proposed as relevant variables for determining the IMF based on theoretical modeling (e.g., \cite{Adams96,Padoan02,Weidner05,Hennebelle13,Fontanot18}) and indirect observational evidences (e.g., \cite{vanDokkum10,MartinNavarro15,Cappellari16,Zhang18,Li23}).

In the present work, we aim to provide constraints on the IMF during the epoch of reionization, i.e., a range of redshift $z\sim$ 6--10 and galaxy properties (e.g., star formation rates $\lesssim 0.1\, M_\odot$ yr$^{-1}$, largely subsolar metallicity, etc.) unexplored so far. Reionization is the process through which the almost neutral intergalactic medium left over from recombination has transitioned again to an ionized state due to the radiation emitted (mainly) by faint primordial galaxies.
{The history of cosmic reionization that is inferred from cosmological data such as the electron scattering optical depth from the \textit{Planck} \linebreak mission \cite{Aghanim20} or other astrophysical probes \cite{Becker13,Becker21,Konno14,McGreer15,Davies18,Mason18,Konno18,Hoag19,Bolan22,Greig22} can set constraints on the level of the ionizing background from faint primordial galaxies. The background level depends mainly on the galaxy number density, on the escape fraction of ionizing photons from primordial galaxies, and on the IMF \cite{Robertson15,Finkelstein19,Lapi22,Gandolfi22}. Given that the number density of faint galaxies in the reionization epoch has been constrained to a good accuracy thanks to HST and JWST surveys \cite{Bouwens21,Oesch18,Bouwens23a,Bouwens23b,Harikane23a,Harikane23b,Harikane22,Donnan23a,Donnan23b,Naidu22,Finkelstein23} (though an extrapolation to faint magnitudes is still required), and that the redshift-dependent escape fraction can be reasonably assessed via simulations \cite{Dayal20,Puchwein19,Faucher20,Mitra23,Kulkarni19a,Katz23} and indirectly inferred from data
\cite{Paardekooper15,Vanzella18,Alavi20,Smith20,Izotov21,Pahl21,Atek22,Naidu22esc,Khaire19,Cain21,Steidel18,Fletcher19}, the reionization observables can become a valuable probe for determining the stellar IMF.}

Technically, we build a semi-empirical model of cosmic reionization by exploiting \linebreak (i) robust measurements of the high-redshift UV luminosity functions for primordial galaxies out to redshift $z\lesssim 12$ from recent JWST data; (ii) data-inferred or simulation-driven assumption on the redshift-dependent escape fraction of ionizing photons from primordial galaxies; (iii) a simple yet flexible parameterization of the IMF; (iv) the \texttt{PARSEC} stellar evolution code to compute the UV and ionizing emission from different stars' masses as a function of age and metallicity; and (v) a few physical constraints related to stellar and galaxy formation in faint galaxies at the reionization redshifts. We then compare our model outcomes with the reionization observables and perform Bayesian inference on the \linebreak IMF parameters.

The paper is structured as follows: in Section \ref{sec|methods}, we describe our methods and analysis; in Section \ref{sec|results}, we present and discuss our results; in Section \ref{sec|summary}, we summarize our findings. The standard $\Lambda$CDM cosmology \cite{Aghanim20} is adopted. % with rounded parameter values: matter density $\Omega_M \approx 0.3$, baryon density $\Omega_b \approx 0.05$, Hubble constant $H_0 = 100\, h$ km s$^{-1}$ Mpc$^{-1}$ with $h\approx 0.7$.

\section{Methods and Analysis}\label{sec|methods}

In this section, we introduce a basic semi-empirical model of reionization, describe the IMF parameterization, discuss galaxy-formation-informed constraints, and present the Bayesian MCMC framework exploited for parameter estimation.

\subsection{Semi-Empirical Model of Reionization}\label{sec|reion}

\begin{figure}[H]
\includegraphics[width=0.775\textwidth]{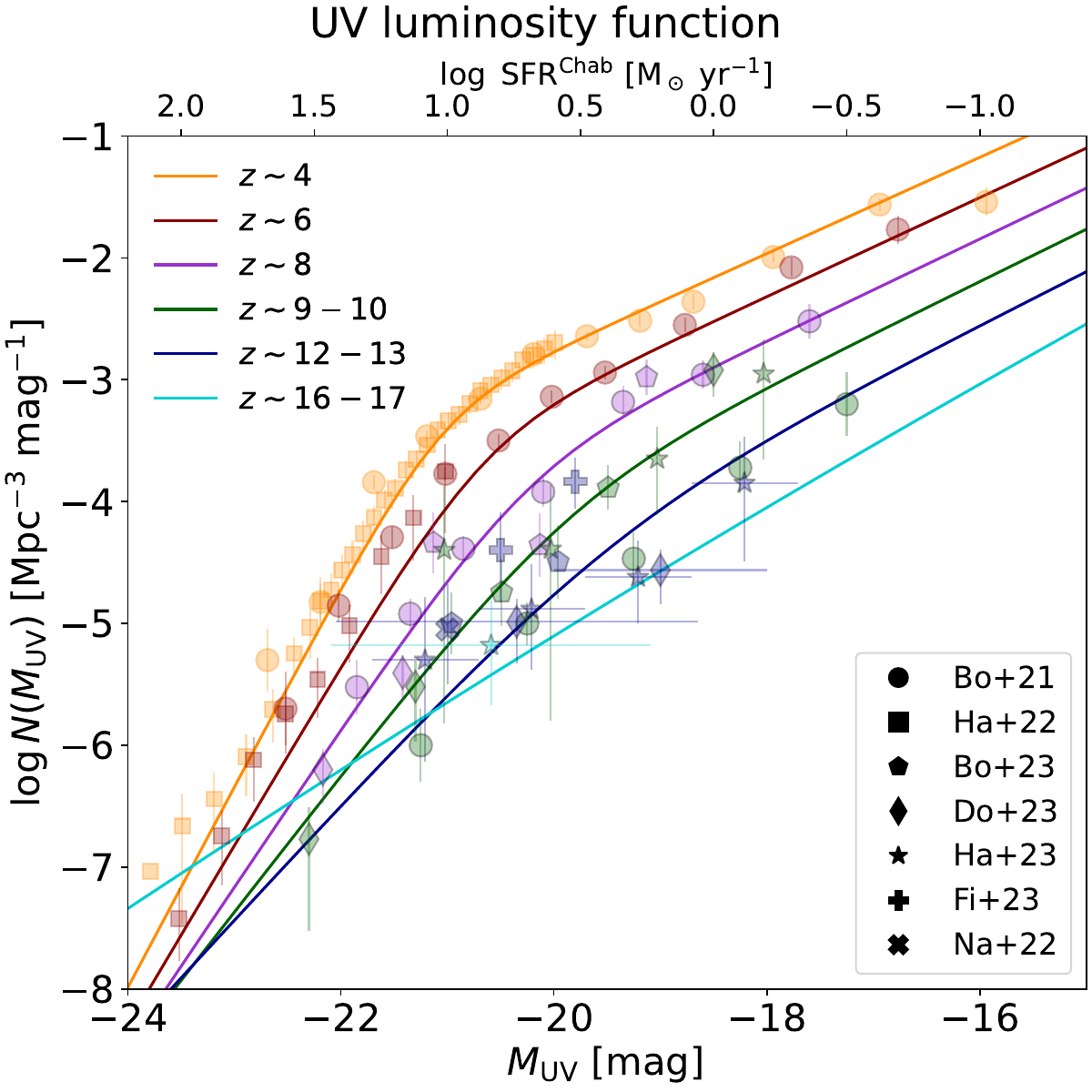}
\caption{The UV luminosity functions at redshifts $z\sim 4$ (orange), $6$ (brown), $8$ (magenta), 9--10 (green), 12--13 (blue), and 16--17 (cyan). Data points are from \cite{Bouwens21,Oesch18} (circles), \cite{Bouwens23a,Bouwens23b} (pentagons), \cite{Harikane23a,Harikane23b} (stars), \cite{Harikane22} (squares), \cite{Donnan23a,Donnan23b} (diamonds), \cite{Naidu22} (crosses), and \cite{Finkelstein23} (pluses). Solid colored lines illustrate a double power law function fitted to the data (see text). The scale on the top axis refers to the SFR on adoption of a Chabrier IMF.}\label{fig|UVLF}
\end{figure}

We aim to construct a simple semi-empirical model of reionization along the lines pursued by \cite{Robertson15,Finkelstein19,Lapi22,Gandolfi22}. To this purpose, we start from the updated determination of the UV luminosity functions calculated by \cite{Bouwens21,Oesch18,Bouwens23a,Bouwens23b,Harikane23a,Harikane23b,Harikane22,Donnan23a,Donnan23b,Naidu22,Finkelstein23} from HST and JWST data out to redshift $z\lesssim 12$. Specifically, in Figure~\ref{fig|UVLF}, we illustrate the binned luminosity functions (filled circles) at $\approx 1600$ {\AA} in the redshift range $z\sim$ 4--16 (color coded). We also display the corresponding double power law rendition (solid lines) in the following form:
\begin{equation}\label{eq|UVLF}
\cfrac{{\rm d}N}{{\rm d}M_{\rm UV}\,{\rm d}V} = \cfrac{\ln(10)}{2.5}\times \cfrac{\phi_\star}{10^{0.4\,(M_{\rm UV}-M_{\rm UV}^\star)\,(\alpha+1)}+10^{0.4\,(M_{\rm UV}-M_{\rm UV}^\star)\,(\beta+1)}}
\end{equation}
where the redshift-dependent parameters $\log \phi_\star\approx -3.146-0.113\, (z-4)$, $M_{\rm UV}^\star\approx -21.081 \linebreak + 0.200\,(z-4)$, $\alpha\approx -1.987-0.014\,(z-4)$, and $\beta\approx -5.116+0.217\,(z-4)$ have been determined by a simple $\chi^2$ fit to the data. Here, $\phi_\star$ plays the role of an overall normalization, $\alpha$ and $\beta$ are the slopes at the faint and bright ends, and $M_{\rm UV}^\star$ is a characteristic magnitude describing the knee of the luminosity function, i.e., the transition between the two power laws. The luminosity function is well determined down to UV magnitude $M_{\rm UV}\approx -17$ for $z\sim 6-10$, which constitutes the redshift range most relevant to our analysis, while the measurements at higher redshifts are more scattered and uncertain. The UV magnitude can be converted into monochromatic UV luminosity at $1600$ {\AA} as $\log L_{\rm UV}$ \mbox{[erg s$^{-1}$ Hz$^{-1}$]} $\approx -0.4\, (M_{\rm UV}-51.6)$.

%\begin{center}
%\end{center}

The intrinsic UV luminosity can be associated with the star formation rate (SFR) as $L_{\rm UV} = \kappa_{\rm UV}\times$ SFR with the UV photons production efficiency $\kappa_{\rm UV}$ depending on the IMF, on the age, and on the chemical composition \cite{Kennicutt12,Madau14,Cai14,Robertson15,Finkelstein19}. For example, a reference value $\kappa_{\rm UV}\approx 1.5\times 10^{28}$ erg s$^{-1}$ Hz$^{-1}$ $\rm{M}_{\odot}^{-1}$ yr applies for a Chabrier IMF, age $\gtrsim 10^8$ years, and sub-solar metallicity. This gives the relation $\log$ SFR [M$_\odot$ yr$^{-1}$] $\approx -0.4\,(M_{\rm UV}+18.5)$, which is adopted to build the scale of the top axis of Figure \ref{fig|UVLF}. {Since most of the ionizing photons are produced in faint galaxies with SFR $\lesssim 1\, M_\odot$ yr$^{-1}$, dust absorption is not an issue, and related corrections of the UV luminosity functions can be neglected in the present \linebreak context \cite{Lapi22,Gandolfi22}.}

The cosmic UV luminosity density is obtained as follows:
\begin{equation}\label{eq|rhoUV}
\rho_{\rm UV}(z)=\int_{-\infty}^{M_{\rm UV}^{\rm lim}}{\rm d}M_{\rm UV}\, \frac{{\rm d}N}{{\rm d}M_{\rm UV}\,{\rm d}V}\, L_{\rm UV}\; ,
\end{equation}
where $M_{\rm UV}^{\rm lim}$ is a limiting magnitude such that, faintward of it, the luminosity function flattens or even bends down because galaxy formation becomes inefficient \cite{Bose18,Romanello21,Munoz22}. The quantity $M_{\rm UV}^{\rm lim}$ is uncertain since the UV luminosity function is observed only for magnitudes brighter than $M_{\rm UV}\approx -17$, but reionization studies indicate that $M_{\rm UV}^{\rm lim}$ must be appreciably fainter than this value \cite{Cai14,Robertson15,Finkelstein19}. In our analysis, it will be treated as a free parameter to be set by comparison with reionization observables. Our results are robust against the detailed shape of the luminosity function faintward of $M_{\rm UV}\sim -16$. For example, instead of extrapolating it with a constant slope $\alpha$ down to $M_{\rm UV}^{\rm lim}$, one can adopt a smooth, progressive bending by multiplying Equation (\ref{eq|UVLF}) by a factor $10^{0.4\, (\alpha+1)/2\times (M_{\rm UV}+16)2/(M_{\rm UV}^{\rm lim}+16)}$ for $M_{\rm UV}\gtrsim -16$, as suggested by \cite{Bouwens22}. We have checked that, in the computation of the UV luminosity density, and hence in all our results, this produces a negligible impact.
%MDPI: We moved the content from footnote into the maintext. Please confirm.
%AUTHORS: OK

\begin{figure}[H]
\includegraphics[width=0.775\textwidth]{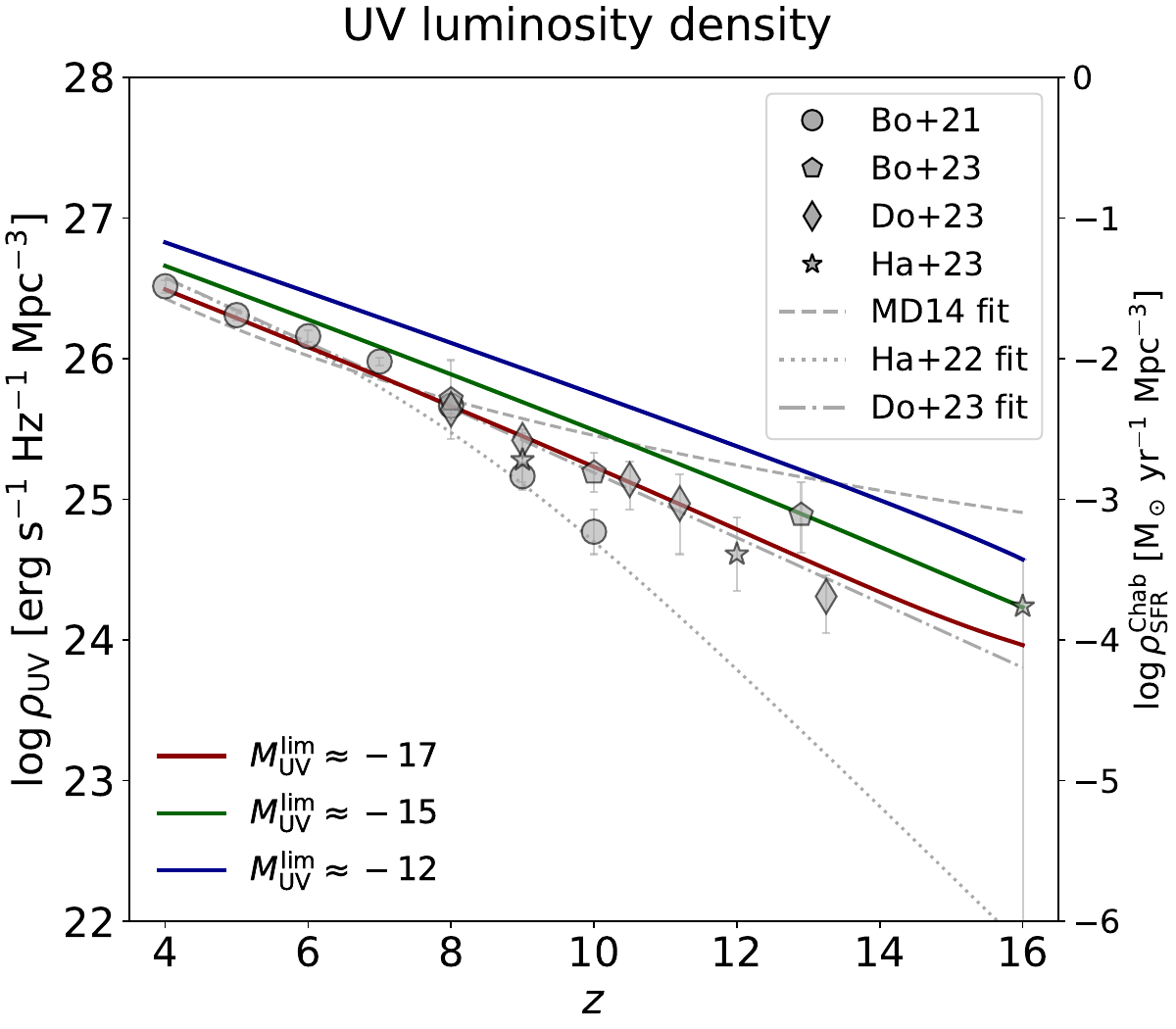}
\caption{The UV luminosity density as a function of redshift. Data points are from \cite{Bouwens21,Oesch18} (circles), \cite{Bouwens23a,Bouwens23b} (pentagons), \cite{Harikane23a,Harikane23b} (stars), \cite{Donnan23a,Donnan23b} (diamonds).
Solid colored lines illustrate the outcomes when integrating the UV luminosity function extrapolated down to magnitude limits $M_{\rm UV}^{\rm lim}\approx -17$ (red), $-15$ (green), and $-12$ (blue). The scale on the right axis refers to the SFR density on adoption of a Chabrier IMF. The gray lines display the parameterization by \cite{Harikane22} (dotted), \cite{Donnan23a} (dot-dashed), \linebreak and \cite{Madau14} (dashed).}\label{fig|rhoUV}
\end{figure}

The resulting UV luminosity density is shown in Figure \ref{fig|rhoUV} for three different values of $M_{\rm UV}^{\rm lim}$. In particular, our result for $M_{\rm UV}\approx -17$ agrees with the recent estimates from JWST data \cite{Bouwens23a,Bouwens23b,Harikane23a,Harikane23b,Donnan23a,Donnan23b}. Note that the latter tends to be appreciably higher for $z\gtrsim 8$ than previous HST estimates \cite{Bouwens21,Oesch18}. This is evident when comparing the fit to JWST data by \cite{Donnan23a} (dot-dashed line) with the HST measurements by \cite{Harikane22} (dotted line). Using a fainter $M_{\rm UV}^{\rm lim}$ enhances the UV luminosity density since this corresponds to {adding} galaxies with lower UV luminosity but much higher number density. The UV luminosity density $\rho_{\rm UV}$ can be expressed as a cosmic SFR density $\rho_{\rm SFR}=\rho_{\rm UV}/\kappa_{\rm UV}$ (i.e., the celebrated `Madau plot') by assuming an IMF to compute $\kappa_{\rm UV}$. In particular, the right axis in Figure \ref{fig|rhoUV} shows the results when adopting a Chabrier IMF. In passing, we note that the classic fitting formula for $\rho_{\rm SFR}(z)$ designed by \cite{Madau14} to describe UV and IR data at $z\lesssim 4$ (dashed line) substantially overestimates the recent JWST determinations for $z\gtrsim 6$ and thus cannot be used in the present context.

%\begin{center}
%\end{center}

The cosmic ionization photon rate is given by the following \cite{Mao07,Cai14,Robertson15,Rutkowski16}:
\begin{equation}\label{eq|Ndotion}
\dot N_{\rm ion}\approx f_{\rm esc}\, k_{\rm ion} \, \cfrac{\rho_{\rm UV}}{\kappa_{\rm UV}} + \dot N_{\rm ion}^{\rm AGN}~,
\end{equation}
where $k_{\rm ion}$ is the production efficiency of ionizing photons (per unit SFR), $f_{\rm esc}$ is the average escape fraction from primordial galaxies, and $\dot N_{\rm ion}^{\rm AGN}$ is the ionization rate from active galactic nuclei (AGNs). It is worth discussing each of these quantities in turn. The ionizing photon production efficiency $k_{\rm ion}$ depends on the adopted IMF, age, metallicity, and other stellar population properties \cite{Mao07,Cai14,Finkelstein19}. For example, a reference often adopted for a Chabrier IMF is $k_{\rm ion}\approx 4\times 10^{53}$ ionizing photons s$^{-1}$ M$_\odot^{-1}$ yr.

\begin{figure}[H]
\includegraphics[width=0.775\textwidth]{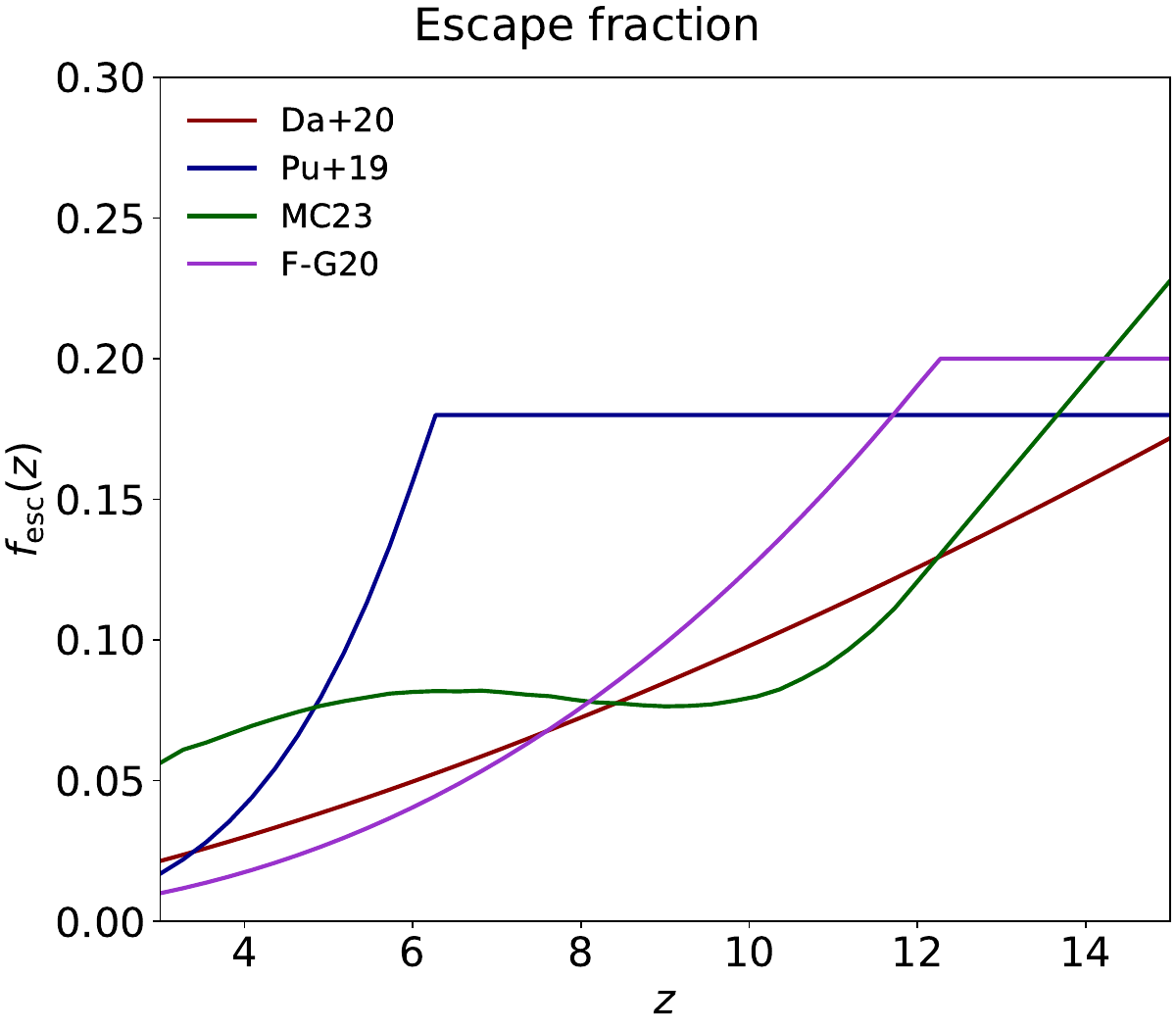}
\caption{The spatially averaged escape fraction of ionizing photons from primordial galaxies as a function of redshift. Colored lines refer to the four different parameterizations by \cite{Dayal20} (red), \cite{Puchwein19} (blue), \cite{Faucher20} (magenta), and \cite{Mitra23} (green).}\label{fig|escape}
\end{figure}

The (spatially averaged) escape fraction $f_{\rm esc}$ from primordial galaxies is still very uncertain, with estimates spanning a range from a few to tens of percent %Please check intended meaning has been retained.
 \cite{Paardekooper15,Vanzella18,Alavi20,Smith20,Izotov21,Pahl21,Atek22,Naidu22esc}. Recent indirect observational estimates of the HI photoionization rate \cite{Finkelstein19,Khaire19} and ionizing photon mean-free path \cite{Cain21}, direct determinations in Lyman-break galaxies \cite{Steidel18,Fletcher19}, and hydrodynamical {radiative transfer} simulations \cite{Kulkarni19a,Katz23} suggest an increasing redshift dependence $f_{\rm esc}(z)$ from a few percent at $z\approx 3$ to about ten percent at $z\sim 10$, albeit with a variety of functional shapes. In this work, we stay agnostic and consider in our analysis the four different escape fraction parameterizations \cite{Dayal20,Faucher20,Puchwein19,Mitra23} that are reported in \linebreak Figure \ref{fig|escape}. Note that the non-monotonic dependence of the green curve originates since this is not a fitting function but actually the result of a non-parametric Gaussian-processes-based reconstruction of $f_{\rm esc}$ by \cite{Mitra23}.

%\begin{center}
%\end{center}

The AGN contribution $\dot N_{\rm ion}^{\rm AGN}$ in Equation (\ref{eq|Ndotion}) can be parameterized as follows \cite{Shen20}:
\begin{equation}\label{eq|Ndotion_AGN}
\dot N_{\rm ion}^{\rm AGN}\approx 1.1\times 10^{50}\,f_{\rm esc}^{\rm AGN} \frac{(1+z)^{5.865}\, e^{0.731\,z}}{15.6+e^{3.055\,z}}\, {\rm photons\;s^{-1}\; Mpc^{-3}}\; ;
\end{equation}

This expression relies on recent estimates of the bolometric AGN luminosity functions, which envisage a paucity of luminous quasars and a minor relevance of faint AGNs for $z\gtrsim 4$, though the debate on the latter point is still open \cite{Shankar07,Giallongo15,Ricci2017,Giallongo19,Kulkarni19b,Ananna20,Shen20}. According to Equation (\ref{eq|Ndotion_AGN}), the AGN contribution to $\dot N_{\rm ion}$ for $z\gtrsim 5$ is negligible with respect to primordial galaxies, even assuming $f_{\rm esc}^{\rm AGN}\sim 100\%$ \cite{Grazian18,Romano19}.

We compute the hydrogen ionizing fraction $Q_{\rm HII}$ as
\begin{equation}\label{eq|QHII}
\dot Q_{\rm HII} = \frac{\dot N_{\rm ion}}{\bar n_{\rm H}}-\frac{Q_{\rm HII}}{
t_{\rm rec}}\; ,
\end{equation}
expressing the interplay between ionization and recombination \cite{Madau99,Loeb01}. {Here, $\bar n_{\rm
H}$ is the mean co-moving hydrogen number density, and $t_{\rm rec}\propto C_{\rm HII}^{-1}$ is the recombination timescale, which inversely depends on the redshift-dependent clumping factor $C_{\rm HII}\approx {\rm min}[1+43\,z^{-1.71},20]$ provided by \cite{Pawlik09,Haardt12}}.
The electron scattering optical depth is given by
\begin{equation}\label{eq|tau_es}
\tau_{\rm es}(<z) = c\, \sigma_{\rm T}\,\bar n_{\rm H}\int^z{\rm d}z'\,f_e\,Q_{\rm HII}(z')
(1+z')^2 \, H^{-1}(z')\; ,
\end{equation}
in terms of the Hubble rate $H(z)=H_0\,[\Omega_M\,(1+z)^3+1-\Omega_M]^{1/2}$ and the number of free electrons $f_e$. More details can be found in \cite{Lapi22}.
%we adopt primordial abundances $Y\approx 0.2454$ and $X\approx 1-Y$, and complete double helium ionization at $z\sim 4$ so that $\eta\approx 2$ for $z\lesssim 4$ and $\eta\approx 1$ for $z\gtrsim 4$.

\subsection{Initial Mass Function}

Since our main aim is to constrain the shape of the IMF, we need an expression of it which is flexible yet does not involve too many parameters. To this purpose, we describe the IMF via a Larson \cite{Larson98} shape $\phi(m_\star) \propto m_\star^{\xi}\, e^{-m_{\star,\rm c}/m_\star}$ in terms of a power law index $\xi<0$ at high star masses and of a characteristic mass $m_{\star,\rm c}$, below which the IMF flattens or even bends downwards. Plainly, the IMF normalization is derived from the condition $\int_{m_{\star,\rm min}}^{m_{\star,\rm max}}{\rm d}m_\star\, m_\star\, \phi(m_\star)=1\, M_\odot$, where $m_{\star,\rm min}\approx 0.08\, M_\odot$ and $m_{\star,\rm max}$ are the minimum and maximum star masses. The upper end is somewhat uncertain; however, for low-metallicity environments, there is observational evidence in local star-forming regions \cite{Schneider18} and from the chemical enrichment in primordial galaxies \cite{Goswami22} which suggests $m_{\star,\rm max}\lesssim 200\, M_\odot$. However, such an outcome depends somewhat on the employed wind mass loss rates in high-mass stellar models in such a way that maximum star masses $m_{\star,\rm max}\lesssim 1000\, M_\odot$ would still be viable.
We adopt an effective upper star mass limit $m_{\star,\rm max}\approx 600\, M_\odot$, but we have checked that the results of this work, which mainly concerns the UV and ionizing emission from massive stars (for which a rather flat luminosity vs. mass relation applies), are marginally impacted by such a value within the range 200--1000 $M_\odot$.% (\hl{see also \linebreak Figure} \ref{fig|stellar}).

\begin{figure}[H]
\includegraphics[width=0.775\textwidth]{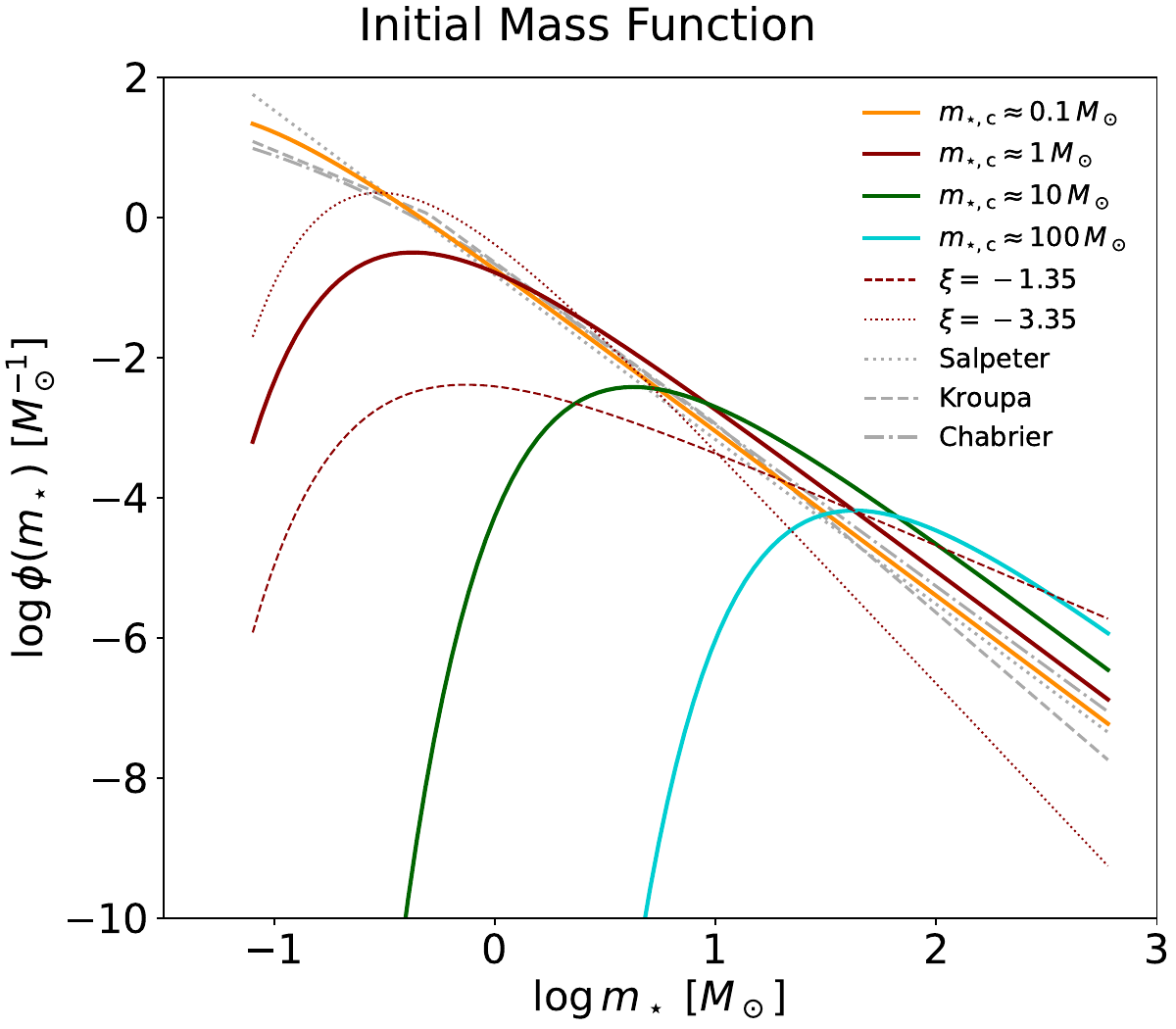}
\caption{The initial mass function (IMF) parameterized in terms of a Larson shape \cite{Larson98}. Solid lines refer to a slope $\xi=-2.35$ and characteristic cutoff masses $m_{\star,\rm c}\approx 0.1\, M_\odot$ (orange), $1\, M_\odot$ (red), $10\, M_\odot$ (green), and $100\, M_\odot$ (cyan). The case with $1\, M_\odot$ is also plotted for different slopes: $\xi=-1.35$ (red dashed) and $\xi=-3.35$ (red dotted). The gray lines correspond to the Salpeter \cite{Salpeter55} (solid), the \linebreak Kroupa \cite{Kroupa01} (dashed), and the Chabrier \cite{Chabrier03} (dot-dashed) IMFs.}\label{fig|imf}
\end{figure}

The shape of the IMF for different values of the parameters $\xi$ and $m_{\star,\rm c}$ is illustrated in Figure \ref{fig|imf}. The index $\xi$ controls the power law behavior for large masses, and hence the relative number of high-mass stars, while the characteristic mass $m_{\star,\rm c}$ determines a smooth cutoff toward small stellar masses. Classic IMF shapes are easily reproduced with this parameterization. The Salpeter \cite{Salpeter55} IMF corresponds to $\xi=-2.35$ and $m_{\star,\rm c}\approx m_{\star,\rm min} = 0.08\, M_\odot$ so that no flattening is present and the IMF is a simple power law. The \linebreak Kroupa \cite{Kroupa01} and Chabrier \cite{Chabrier03} IMFs correspond to $\xi\approx -2.7$ or $-2.3$ and $m_{\star,\rm c}\approx 0.15\, M_\odot$ so that the slope of the IMF progressively flattens below $1\, M_\odot$, although no real cutoff is enforced. A top-heavy IMF can be obtained either with flat power law shapes and no cutoff so as to enhance the production of very massive stars or with a more canonical power law index and a cutoff below $1\, M_\odot$ that strongly suppresses the formation of small/intermediate-mass stars.

%\begin{center}
%\end{center}
%MDPI: We swapped the position of Figures 4 and 5 to make the citations appear in numerical order. Please confirm.
% AUTHORS: No, from a scientific point of view it is better to keep the previous version. We have reverted to it and removed a anticipated citation of the figure so that now the numerical ordering is consistent.

\begin{figure}[H]
%\includegraphics[width=0.575\textwidth]{kappa_UV.pdf}\\
%\\
%\includegraphics[width=0.575\textwidth]{kappa_ion.pdf}\\
%\\
%\includegraphics[width=0.575\textwidth]{MoverL.pdf}\\
\includegraphics[width=0.775\textwidth]{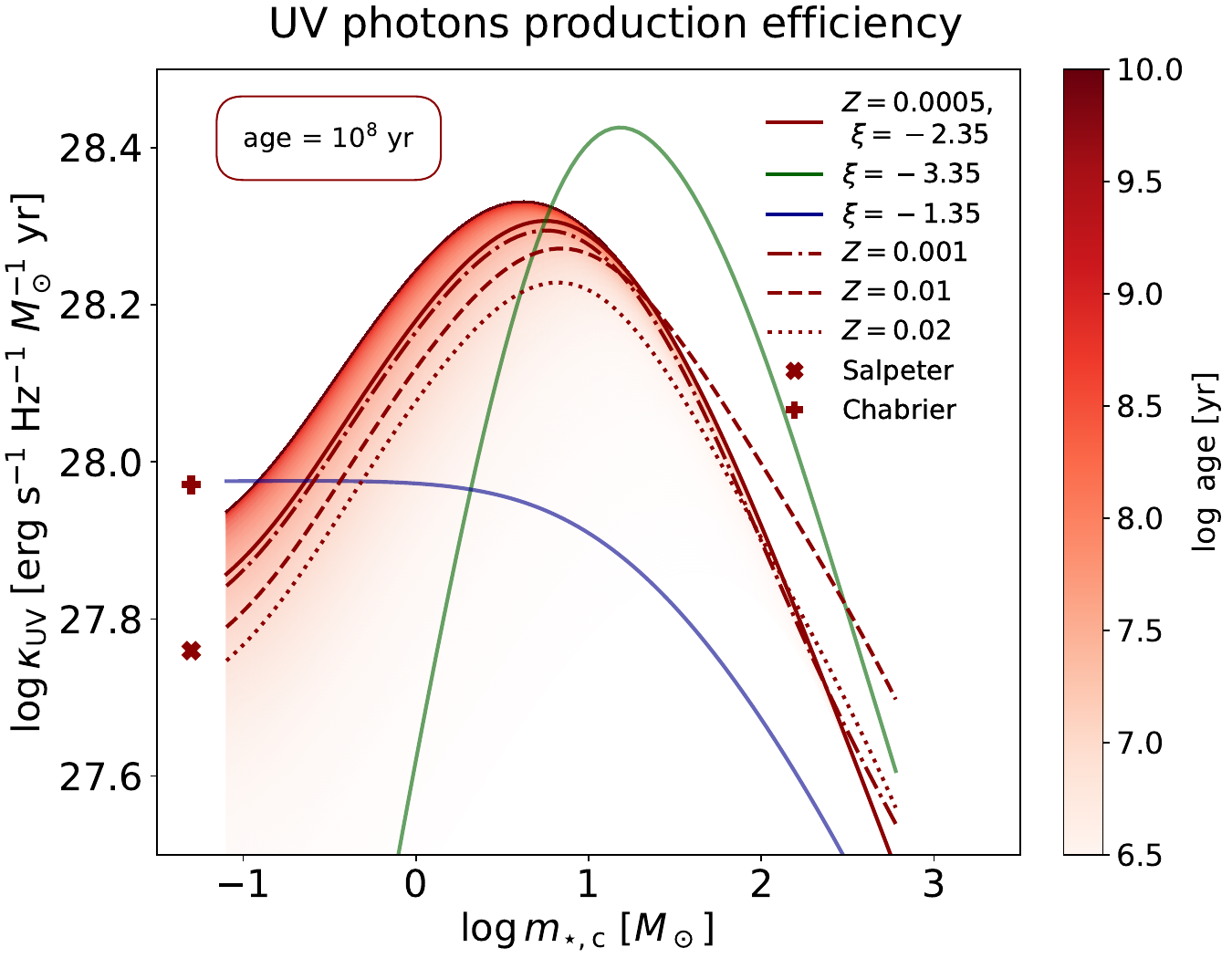}\\
\\
\includegraphics[width=0.775\textwidth]{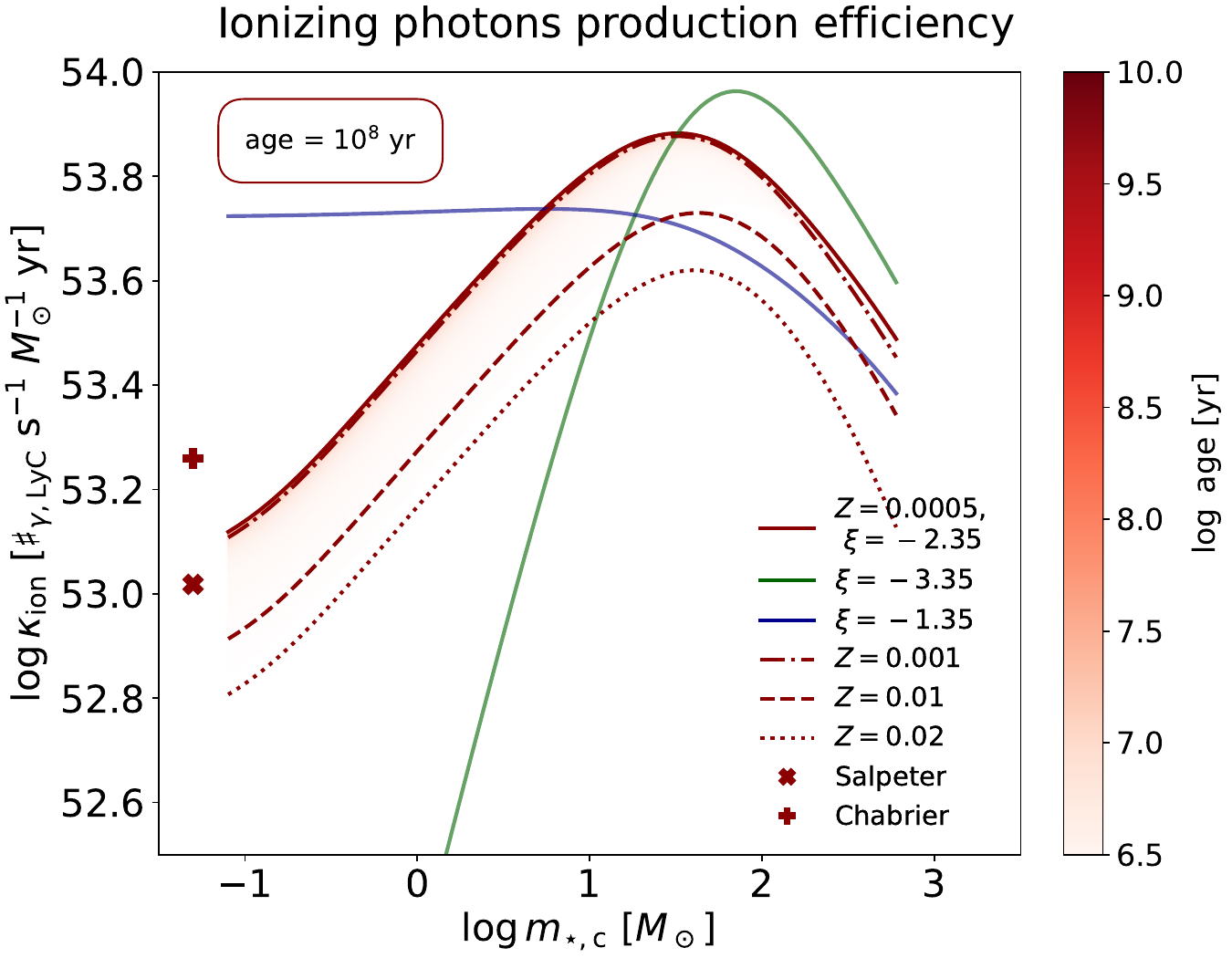}\\
\caption{The production efficiency of UV photons (top panel) and ionizing photons (bottom panel) as a function of the characteristic mass of the IMF. In both panels, the colored lines refer to an age of \linebreak $100$ Myr. The solid line is for an IMF slope $\xi=-2.35$ and a metallicity $Z=0.0005$. At fixed $Z=0.0005$, the green line is for $\xi=-3.35$ and the blue for $\xi=-1.35$. At fixed $\xi=-2.35$, the dot-dashed line is for $Z=0.001$, the dashed for $Z=0.01$, and the dotted for $Z=0.02$. The plus and cross display the reference values for the Salpeter and the Chabrier IMFs. Finally, at fixed $\xi=-2.35$ and $Z=0.0005$, the shaded area shows the effect of changing the age from $3$ Myr to $10$ Gyr.}\label{fig|stellar}
\end{figure}

The quantities $k_{\rm UV}$ and $k_{\rm ion}$ entered into Equation (\ref{eq|Ndotion}) crucially depend on the adopted IMF. We compute them by exploiting the \texttt{PARSEC} \cite{Bressan12,Goswami22} stellar evolutionary code, in particular, its recent extension to high-mass stars \cite{Goswami22}. The individual stellar tracks are used to compute the mean production rate of UV and ionizing photons, and then $\kappa_{\rm UV}$ and $\kappa_{\rm ion}$ are obtained by appropriately averaging these over the IMF. The outcome is illustrated in Figure \ref{fig|stellar}, where $k_{\rm UV}$ and $k_{\rm ion}$ are plotted as a function of the characteristic mass of the IMF for different values of age, metallicity, and IMF slope.

The red solid line refers to a very low metal content $Z=0.0005$, age of $100$ Myr, and standard IMF slope $\xi=-2.35$. Increasing $m_{\star,\rm c}$ removes progressively star masses smaller than this value from the IMF, hence, weighting more intermediate- and/or high-mass stars. Therefore, the $k_{\rm UV}$ and $k_{\rm ion}$ follow a bell-shaped trend as a function of $m_{\star,\rm c}$ since they are maximized when $m_{\star,\rm c}$ has a value around the star masses, which are the most important contributors to the production of UV or ionizing photons. For $k_{\rm UV}$, this occurs at around a few $M_\odot$, while, for $k_{\rm ion}$, it is at around a few tens of $M_\odot$.

The red dot-dashed, dashed, and dotted curves show the effect of increasing the metallicity. This slightly shifts the peak of the curve toward higher $m_{\star,\rm c}$ and tends to appreciably reduce the production efficiency, especially in the case of ionizing photons. The red shaded area (as labeled in the color bar) displays the impact of the stellar population age. $k_{\rm ion}$ is mainly contributed by massive stars with very low age so it saturates after a few $10^7$ of years. %Please check intended meaning has been retained here and in similar instances that follow.
 $k_{\rm UV}$ is contributed also by intermediate-mass stars so it requires several $10^8$ of years to saturate. In the present context of faint galaxies in the reionization era, we can confidently exploit the results for low metallicities $Z\lesssim 0.002$ and age $\lesssim$ a few $10^8$ years.

The blue and green solid lines show, instead, the outcomes when varying the IMF slope. Flattening (steepening) the IMF slope smooths (exacerbates) the dependence of $k_{\rm UV}$ and $k_{\rm ion}$ on the IMF characteristic mass and decreases (increases) the maximal production efficiencies. This is particularly evident for $k_{\rm UV}$ since, e.g., adopting a very flat slope $\xi=-1.35$ greatly under-weights the intermediate-mass stars that are significant contributors to the total UV photon production.

%\begin{center}
%\end{center}

\subsection{Constraints from Galaxy Formation}\label{sec|abma}

We now introduce three constraints related to galaxy and star formation in primordial galaxies that will help to break the degeneracies among, and hence better determine, the fitting parameters.

The first constraint concerns the link between the limiting UV magnitude $M_{\rm UV}^{\rm lim}$ and the minimum halo mass for galaxy formation $M_{\rm H}^{\rm GF}$. {The rationale is that, for halo masses smaller than $M_{\rm H}^{\rm GF}$, galaxy formation is offset by various astrophysical processes \cite{Efstathiou92,Sobacchi13,Cai14,Finkelstein19}. This, in turn, must set $M_{\rm UV}^{\rm lim}$ since, at fainter magnitude, the UV luminosity function should not rise any longer, i.e., the number of faint galaxies must be appreciably reduced.}
%More specifically, below the critical halo mass $M_{\rm H}^{\rm GF}$ the galaxy formation efficiency may be reduced : molecular cooling may be hindered and atomic cooling may be limited given the low metallicities expected at high redshift; SN feedback can easily quench star formation in low-mass halos; star formation may be photo-suppressed by the intense diffuse UV background; the formation of massive stars at low metallicities may originate additional radiative feedback processes; etc.
%Such a threshold value $M_{\rm H}^{\rm GF}$ is often taken around a few $10^8$ M$_\odot$ with a possible (weak) redshift dependence. %Remarkably, it has also been pointed out that this can alleviate the missing satellite problem, because the number density of small mass halos where galaxy formation can take place becomes closer to the number of visible satellites in the Milky Way \cite{Bullock17}.
We conservatively assume $M_{\rm H}^{\rm GF}\approx 3\times 10^8$ M$_\odot$, which may be related to photo-suppression of star formation by the UV background \cite{Finkelstein19}.

\begin{figure}[H]
\includegraphics[width=0.775\textwidth]{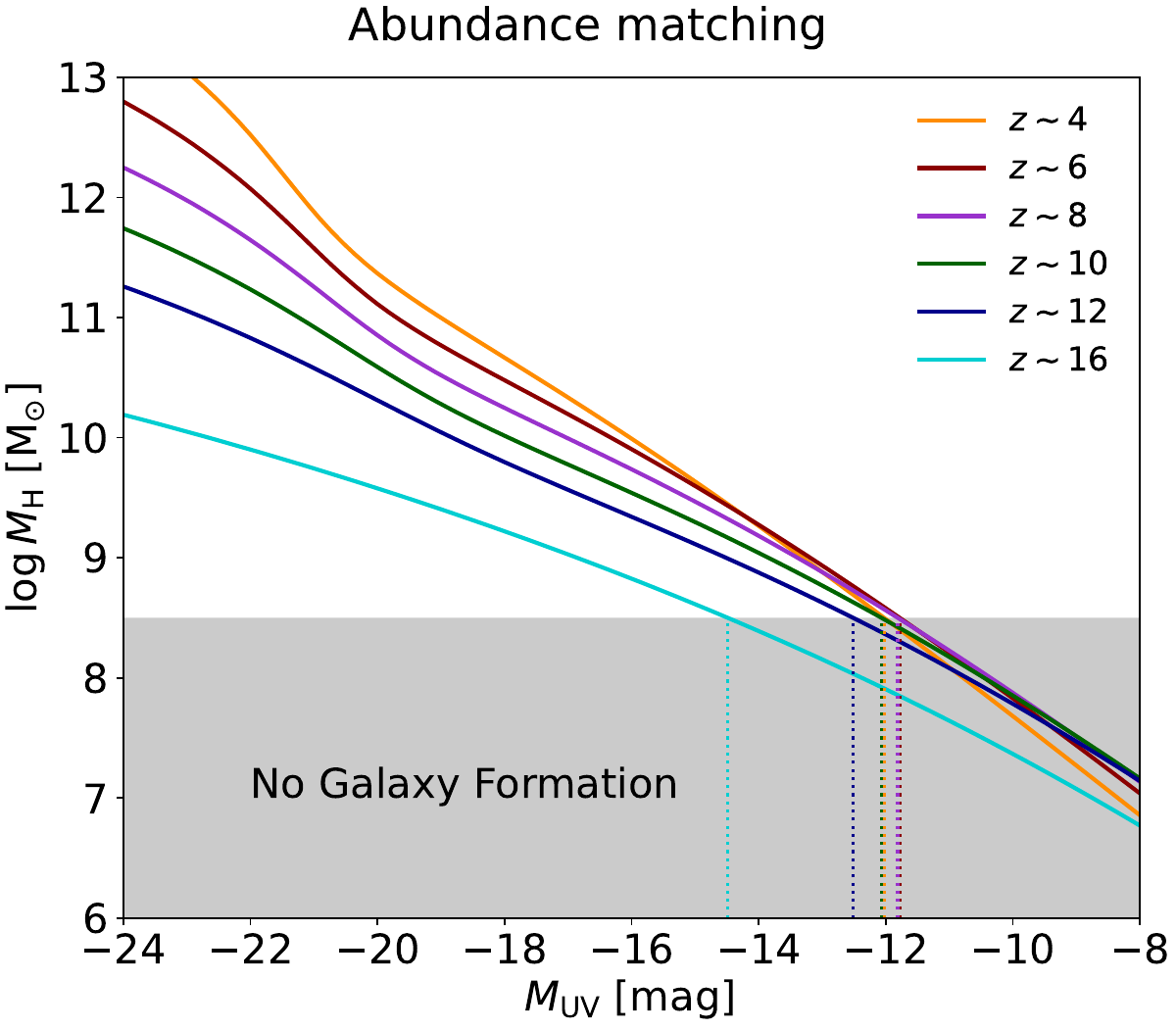}
\caption{The abundance matching relationship between the halo mass and the UV magnitude. Colored lines refer to different redshifts: $z\sim 4$ (orange), $6$ (brown), $8$ (magenta), $10$ (green), $12$ (blue), and $16$ (cyan). The shaded gray area illustrates the regions where galaxy formation is hindered (see text), corresponding to the UV magnitude limits highlighted by the dotted vertical lines.}\label{fig|abma}
\end{figure}

To associate the UV magnitude and the host halo mass, we exploit an abundance matching technique \cite{Aversa15,Moster18,Cristofari19,Behroozi20}, i.e., we require the cumulative number densities in galaxies and halos to match as follows:
\begin{equation}\label{eq|abma}
\int_{M_{\rm H}}^{+\infty}{\rm d}M_{\rm H}'\;\frac{{\rm d}N}{{\rm d}M_{\rm H}'\, {\rm d}V}(M_{\rm H}',z) = \int_{-\infty}^{M_{\rm UV}}{\rm d}M_{\rm UV}'\;\frac{{\rm d}N}{{\rm d}M_{\rm UV}'\, {\rm d}V}(M_{\rm UV}',z)
\end{equation}
which yields a relation $M_{\rm H}(M_{\rm UV},z)$. The integrand on the left-hand side is the CDM halo mass function by \cite{Tinker08} computed via the \texttt{COLOSSUS} package \cite{Diemer18}. {The abundance matching outcome is plotted in Figure \ref{fig|abma} for different redshifts (color coded) in the range relevant for reionization with the region below $M_{\rm H}^{\rm GF}$ highlighted in gray shades.}

%\begin{center}
%\end{center}

Thus, we now require
\begin{equation}\label{eq|MnoGF}
M_{\rm H}(M_{\rm UV}^{\rm lim},z) \approx M_{\rm H}^{\rm GF}\;,
\end{equation}
meaning that, if galaxy formation is offset for halo masses $M_{\rm H}\lesssim M_{\rm H}^{\rm GF}$, the UV luminosity function should flatten or even bend down for the correspondingly faint UV magnitudes $M_{\rm UV}\gtrsim M_{\rm UV}^{\rm lim}$. For redshifts $z\sim$ 4--12, the resulting values of $M_{\rm UV}^{\rm lim}$ turn out to be around $-12$. Note that, in our Bayesian analysis below, we adopt a $0.25$ dex scatter around \linebreak Equation (\ref{eq|MnoGF}) mainly related to uncertainties in the value $M_{\rm H}^{\rm GF}$ (see~\cite{Cai14,Finkelstein19}).

\begin{figure}[H]
\includegraphics[width=0.775\textwidth]{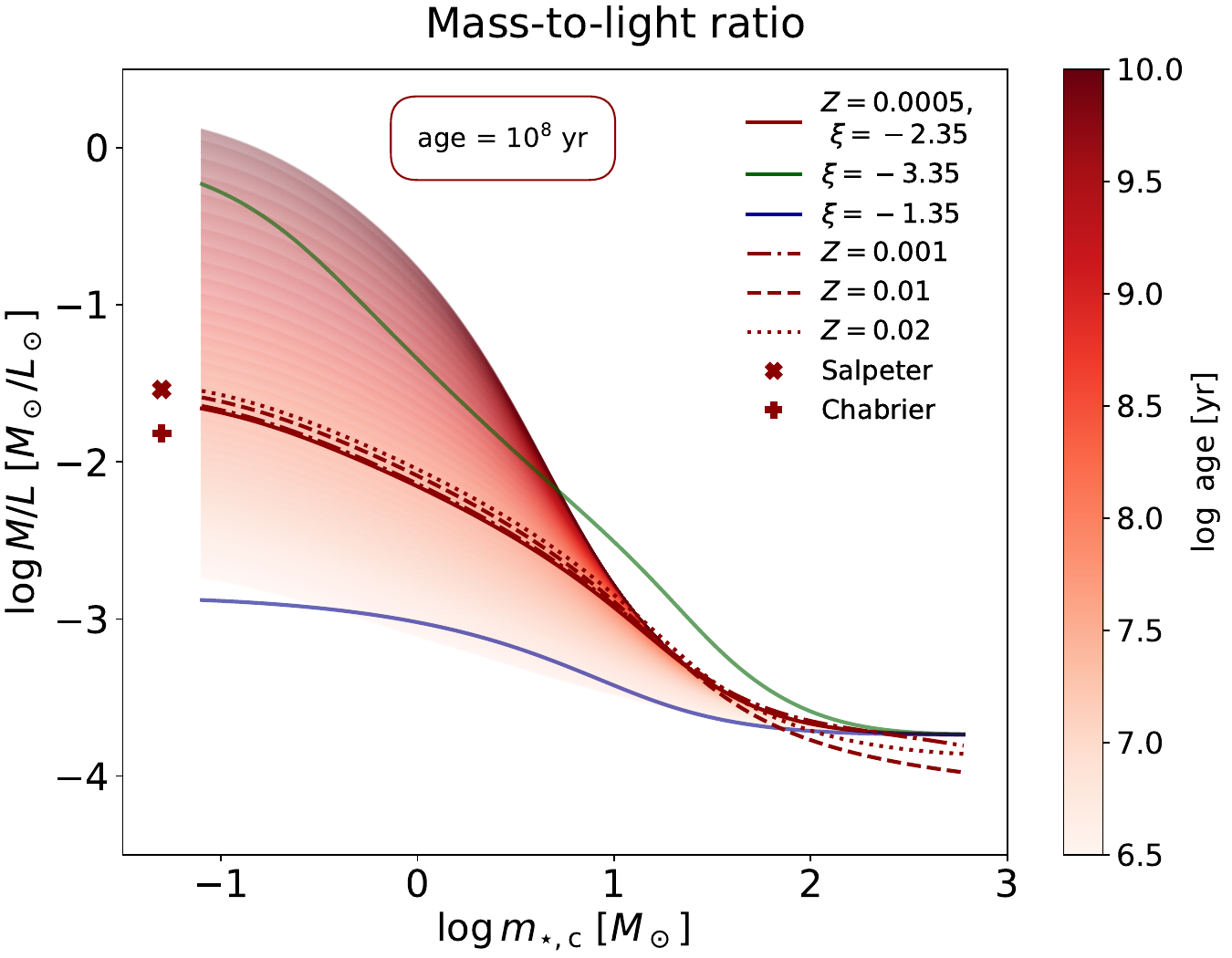}\\
\caption{The average $\langle M_\star/L_{\rm UV}\rangle$ as a function of the characteristic mass of the IMF. The colored lines refer to an age of $100$ Myr. The solid line is for an IMF slope $\xi=-2.35$ and a metallicity $Z=0.0005$. At fixed $Z=0.0005$, the green line is for $\xi=-3.35$ and the blue for $\xi=-1.35$. At fixed $\xi=-2.35$, the dot-dashed line is for $Z=0.001$, the dashed for $Z=0.01$, and the dotted for $Z=0.02$. The plus and cross display the reference values for the Salpeter and the Chabrier IMFs. Finally, at fixed $\xi=-2.35$ and $Z=0.0005$, the shaded area shows the effect of changing the age from $3$ Myr to $10$ Gyr.}\label{fig|MoverL}
\end{figure}

The second constraint involves the efficiency of galaxy formation expected in small halos. We can compute a cosmic-averaged star formation efficiency $\bar\epsilon_\star$ as the ratio between the co-moving density in stars and as a total number of baryons (integrated down to $M_{\rm UV}^{\rm lim}$ and down to the corresponding $M_{\rm H}^{\rm GF}$, respectively) as follows:
\begin{equation}\label{eq|efficiency}
\bar\epsilon_\star(z)\approx \int_{-\infty}^{M_{\rm UV}^{\rm lim}}{\rm d}M_{\rm UV}\, \frac{{\rm d}N}{{\rm d}M_{\rm UV}\,{\rm d}V}\, L_{\rm UV}\times \langle \cfrac{M_\star}{L_{\rm UV}}\rangle \Bigg{/} \int_{M_{\rm H}^{\rm GF}}^\infty{\rm d}M_{\rm H}\, f_b\,M_{\rm H}\, \cfrac{{\rm d}N}{{\rm d}M_{\rm H}\,{\rm d}V}
\end{equation}
where $f_b\approx 0.16$ is the cosmic baryon fraction, and $\langle M_\star/L_{\rm UV}\rangle$ is the IMF-averaged stellar mass-to-light ratio. The latter quantity can be computed via the \texttt{PARSEC} stellar evolution tracks after proper weighting over the adopted IMF and is displayed in Figure \ref{fig|MoverL} as a function of the IMF parameters, metallicity, and age. The dependence on age is easy to understand; the mass-to-light ratio increases in older stellar populations since the UV luminosity from massive, short-living stars decreases as they progressively die, while the stellar mass is mostly contributed by the numerous low-mass, long-living stars. At a fixed age, increasing the IMF characteristic mass or flattening the IMF slope makes the $\langle M_\star/L_{\rm UV}\rangle$ decrease since low-mass stars are under-weighted. The dependence on metallicity is very weak.

%\begin{center}
%\end{center}

Constraints on $\bar \epsilon_\star$ come from abundance matching arguments involving the halo and the stellar mass functions \cite{Moster13,Aversa15}. In the local Universe, these are also consistent with independent observational determinations of the stellar mass vs. halo mass relation from satellite kinematics \cite{Lange19}, galaxy rotation curves \cite{Lapi18}, and weak lensing \cite{Mandelbaum16}. In the present context, we are interested in a range of redshift $z\sim 6-10$ and halo masses $M_{\rm H}\lesssim 10^9\, M_\odot$, where the abundance matching results must be somewhat extrapolated. However, the shape and evolution of the stellar mass vs. halo mass relation suggest that an upper limit $\bar\epsilon_{\star,\rm max}\approx 0.1$ can be safely adopted in small halos, though we allow for a rather large $1\sigma$ dispersion of $0.3$ dex around this value.

Finally, a third relevant constraint involves the characteristic mass $m_{\star,\rm c}$ of the IMF. During the reionization era at around $z\sim 6-10$, the CMB temperature can be as high as $T_{\rm CMB}(z)\gtrsim 20$ K. This provides a basic thermal background that can limit the fragmentation of gas clouds into small cores and the ensuing condensation into low-mass stars. A lower limit for the star masses can be obtained by considering the thermal Jeans mass as follows \cite{Padoan02,Hennebelle13}:
\begin{equation}\label{eq|MJeans}
m_{\star,\rm J}(z)  \approx 1.2\times \left[\cfrac{T_{\rm CMB}\,(1+z)}{10\,{\rm K}}\right]^{3/2}\, M_\odot
\end{equation}
where $T_{\rm CMB}\approx 2.73$ K is the present CMB temperature (a standard mean density of \linebreak$10^3$ cm$^{-3}$ is adopted). Note that such a value is a conservative lower limit in the sense that turbulent or magnetic pressure can lead to even larger values. We allow for a rather large $1\sigma$ dispersion of $0.25$ dex around such a value.

\subsection{Bayesian Analysis}\label{sec|Bayes}

The discussion provided so far highlights that three parameters are relevant in our framework: the limiting UV magnitude $M_{\rm UV}^{\rm lim}$, the IMF characteristic mass $m_{\star, \rm c}$, and the IMF slope $\xi$. We estimate these parameters via a Bayesian MCMC framework. Specifically, {we adopt a standard log likelihood} $\ln \mathcal{L_{\rm data}}(\theta)\equiv -\sum_i \chi_i^2(\theta)/2$ in terms of $\theta=\{M_{\rm UV}^{\rm lim},\log m_{\star, \rm c},\xi\}$, where the sum is over different observables. For each of these, the $\chi_i^2= \sum_j [\mathcal{M}(z_j,\theta)-\mathcal{D}(z_j)]^2/\sigma_{\mathcal{D}}^2(z_j)$ is obtained in terms of the model $\mathcal{M}(z_j,\theta)$ and the data values $\mathcal{D}(z_j)$ with their uncertainties $\sigma_{\mathcal{D}}^2(z_j)$, summing over different redshifts $z_j$ (when applicable). {Our overall data sample is fully reported in Table \ref{tab|data} and is constituted by observational measurements of} the ionizing photon rate \cite{Becker13,Becker21}, the volume filling factor of ionized hydrogen \cite{Konno14,McGreer15,Davies18,Mason18,Konno18,Hoag19,Bolan22,Greig22}, and the electron scattering optical depth \cite{Aghanim20}.

\begin{table}[H]
\caption{Data considered for the Bayesian analysis of this work (see text for details).%, referring to the ionizing photon rate $\dot N_{\rm ion}$, volume filling fraction $Q_{\rm HII}$ of ionized hydrogen, optical depth $\tau_{\rm es}$ for electron scattering.
}\label{tab|data}
\tablesize{\footnotesize}
\newcolumntype{H}{>{\centering\arraybackslash}X}
\begin{tabularx}{\textwidth}{p{3cm}<{\centering} Cp{2.5cm}<{\centering}CC}
\toprule
\textbf{Observable [Units]} & \textbf{Redshifts} & \textbf{Values} & \textbf{Uncertainty} & \textbf{Ref.}\\
\midrule
&&&&\\
$\log \dot N_{\rm ion}$ $[$ s$^{-1}$ Mpc$^{-3}]$&&&&\\
&&&&\\
& $\{4.0,4.8\}$ & $\{50.86,50.99\}$ & $\{0.39,0.39\}$ & \cite{Becker13} \\
&&&&\\
& $\{5.1\}$ & $\{51.00\}$ & $\{0.15\}$ & \cite{Becker21} \\
&&&&\\
$Q_{\rm HII}$&&&&\\
&&&&\\
& $\{7.0\}$ & $\{0.41\}$ & $\{0.13\}$ & \cite{Mason18} \\
&&&&\\
& $\{7.6\}$ & $\{0.12\}$ & $\{0.07\}$ & \cite{Hoag19} \\
&&&&\\
& $\{6.6,6.9,7.3\}$ & $\{0.30,0.50,0.55\}$ & $\{0.20,0.10,0.25\}$ & \cite{Konno14,Konno18} \\
&&&&\\
& $\{7.6\}$ & $\{0.83\}$ & $\{0.10\}$ & \cite{Bolan22} \\
&&&&\\
& $\{7.3\}$ & $\{0.49\}$ & $\{0.11\}$ & \cite{Greig22} \\
&&&&\\
& $\{7.1,7.5\}$ & $\{0.48,0.60\}$ & $\{0.26,0.22\}$ & \cite{Davies18} \\
&&&&\\
& $\{5.6,5.9\}$ & $\{<0.04,<0.06\}$ & $\{0.05,0.05\}$ & \cite{McGreer15} \\
&&&&\\
$\tau_{\rm es}$&&&&\\
&&&&\\
& $\{\infty\}$ & $\{0.054\}$ & $\{0.007\}$ & \cite{Aghanim20} \\
&&&&\\
\bottomrule
\end{tabularx}
\end{table}

{We also include in the overall log likelihood $\ln \mathcal{L_{\rm tot}}=\ln \mathcal{L_{\rm data}} + \ln \mathcal{L_{\rm GF}}+\ln \mathcal{L_\star}+\ln \mathcal{L_{\rm J}}$ the galaxy-formation-informed constraints discussed in Section \ref{sec|abma}}: $\ln \mathcal{L_{\rm GF}}\sim -\sum_j\,[\log M_{\rm H}\linebreak(M_{\rm UV}^{\rm lim},z_j) - \log M_{\rm H}^{\rm GF}]^2/2\,\sigma_{\rm GF}^2$ with $\log M_{\rm H}^{\rm GF}\,[M_\odot]\approx 8.5$ and $\sigma_{\rm GF}\approx 0.25$ dex, {to implement the relation between the limiting UV luminosity and minimum halo mass from \linebreak Equation (\ref{eq|MnoGF})}; $\ln \mathcal{L}_{\star} \sim \sum_j\,\ln \{\sqrt{\pi/2}\, \sigma_{\star}\, [1+{\rm erf}(\log \bar\epsilon_{\star,\rm max}-\log \bar\epsilon_{\star}(z_j))/\sqrt{2}\,\sigma_{\star}]\}$ with \linebreak$\bar\epsilon_{\star,\rm max}\approx 0.1$ and $\sigma_{\star}\approx 0.3$ dex, {to implement the upper bound on the cosmic-averaged star formation efficiency $\bar\epsilon_\star$ from Equation (\ref{eq|efficiency})}; and
$\ln \mathcal{L}_{\rm J} \sim \sum_j\,\ln \{\sqrt{\pi/2}\, \sigma_{\rm J}\, [1+{\rm erf}(\log m_{\star,\rm c}\linebreak-\log m_{\star,\rm J}(z_j))/\sqrt{2}\,\sigma_{\rm J}]\}$ with $\sigma_{\rm J}\approx 0.25$ dex, {to implement the lower bound on the \linebreak characteristic mass of the IMF as provided by the thermal Jeans mass $m_{\star,\rm J}$ from \linebreak Equation (\ref{eq|MJeans}).}

We adopt flat priors $\pi(\theta)$ on the parameters $M_{\rm UV}^{\rm lim}\in [-18,-8]$, $\log m_{\star,\rm c}\,[M_\odot]\in [-1,3]$, and $\xi\in [-5,0]$ and then sample the parameter posterior distributions $\mathcal{P}(\theta) \propto \mathcal{L}_{\rm tot}(\theta)\,\pi(\theta)$ via the MCMC Python package \texttt{emcee} \cite{Foreman13}.

%, running it with $10^4$ steps and $100$ walkers; each walker is initialized with a random position extracted from the priors discussed above. To speed up convergence, we adopt a mixture of differential evolution and snooker moves of the walkers, in proportion of $0.8$ and $0.2$ respectively, that emulates a parallel tempering algorithm. After checking the auto-correlation time, we remove the first $30\%$ of the flattened chain to ensure burn-in; the typical acceptance fractions of the various runs are around $30\%$.

\section{Results and Discussion}\label{sec|results}

\begin{figure}[H]
\includegraphics[width=0.775\textwidth]{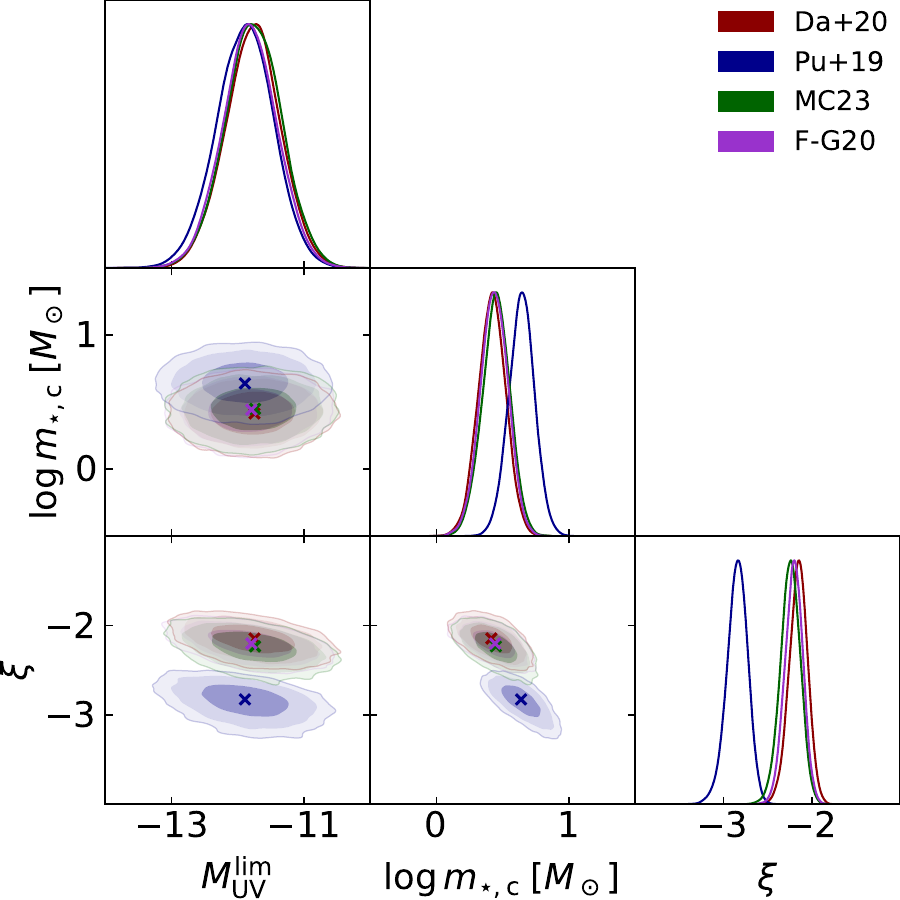}
\caption{MCMC posterior distributions of the limiting UV magnitude $M_{\rm UV}^{\rm lim}$, on the characteristic cutoff mass of the IMF $m_{\star,\rm c}$, and on the IMF slope $\xi$. Colored contours/lines refer to each of the escape fraction parameterizations by \cite{Dayal20} (red), \cite{Puchwein19} (blue), \cite{Mitra23} (green), and \cite{Faucher20} (magenta). The marginalized distributions are in arbitrary units (normalized to 1 at their maximum value).}\label{fig|MCMC}
\end{figure}

The results of our Bayesian analysis are displayed in Figure \ref{fig|MCMC} and reported in \linebreak Table \ref{tab|results}. Specifically, in Figure \ref{fig|MCMC}, we illustrate the MCMC posterior distributions on the UV limiting magnitude $M_{\rm UV}^{\rm lim}$, the IMF characteristic mass $m_{\star,\rm c}$, and the IMF slope $\xi$ for the four different escape fraction prescriptions considered in this work (color coded). The crosses mark the bestfit values of the parameters, and the marginalized distributions are normalized to unity at their maximum. In Table \ref{tab|results}, we summarize the marginalized posterior estimates of the parameters (median values, $1\sigma$ credible intervals, and bestfit values) together with the corresponding reduced $\chi_r^2$ of the fits.

%\begin{center}
%\end{center}
\vspace{-6pt}
\begin{table}[H]\label{tab|results}
\caption{Marginalized posterior estimates (median and $1\sigma$ confidence intervals and bestfit in square brackets) of the parameters from the MCMC analysis for the different escape fraction parameterizations (see text for details). Columns report the values of the limiting UV magnitude $M_{\rm UV}^{\rm lim}$, the characteristic cutoff mass $m_{\star,\rm c}$ of the IMF, the IMF slope $\xi$, and the reduced $\chi_r^2$ for the \linebreak overall fit.}\label{tab|results}
\newcolumntype{C}{>{\centering\arraybackslash}X}
\begin{tabularx}{\textwidth}{lCCCC}
\toprule
\boldmath{$f_{\rm esc}(z)$} &\boldmath{$M_{\rm UV}^{\rm lim}$} & \boldmath{$\log m_{\star,\rm c}\, [M_\odot]$} & \boldmath{$\xi$} & \textbf{$\chi_r^2$} \\
\midrule
&&&&\\
Da+20 \cite{Dayal20}  &  $-11.77^{+0.42}_{-0.42}$ [$-$11.75] & $0.42^{+0.11}_{-0.11}$ [0.42] & $-2.15^{+0.11}_{-0.11}$ [$-$2.15] & $1.63$\\
&&&&\\
Pu+19 \cite{Puchwein19}  &  $-11.89^{+0.44}_{-0.44}$ [$-$11.89] & $0.64^{+0.09}_{-0.09}$ [0.64] & $-2.84^{+0.13}_{-0.13}$ [$-$2.83] & $1.31$\\
&&&&\\
F-G20 \cite{Faucher20}  &  $-11.82^{+0.42}_{-0.42}$ [$-$11.79] & $0.44^{+0.11}_{-0.11}$ [0.44] & $-2.20^{+0.11}_{-0.09}$ [$-$2.20] & $2.32$\\
&&&&\\
MC23 \cite{Mitra23}  &  $-11.76^{+0.43}_{-0.43}$ [$-$11.74] & $0.45^{+0.11}_{-0.11}$ [0.45] & $-2.24^{+0.12}_{-0.11}$ [$-$2.23] & $1.13$\\
&&&&\\
\bottomrule
\end{tabularx}
\end{table}

Pleasingly, the results for different escape fraction prescriptions are quite stable. This is due to the similarities of the escape fraction models $f_{\rm esc}(z)$ illustrated in Figure \ref{fig|escape} in the redshift range $z\sim$ 6--10 where reionization data are available. Not surprisingly, the most variant results are for the escape fraction parameterization by \cite{Puchwein19}, which implies that $f_{\rm esc}$ saturates to high values $f_{\rm esc}\approx 0.2$ at $z\gtrsim 6$, while the other escape fraction models envisage $f_{\rm esc}\lesssim 10\%$ at those redshifts.

The bestfit values of the fitting parameter read $M_{\rm UV}^{\rm lim}\approx -12$ for the UV limiting magnitude, $m_{\star,\rm c}\approx$ a few $M_\odot$ for the IMF characteristic mass, and $\xi\approx -2.2$ for the IMF slope. These are mainly determined by the need to have enough UV and ionizing photons to reproduce the reionization data but still meet the galaxy-formation-informed constraints described in Section \ref{sec|abma}.
Remarkably, the IMF slope is, within uncertainties, consistent with the standard value $\xi\approx -2.3$ of a Salpeter shape. Top-heavy IMFs with a flat slope $\xi\gtrsim -1$ are excluded at more than $5\sigma$. This is because such slopes tend to largely overproduce ionizing photons, to the point of inducing a very early reionization history. The effects can be partly compensated only by changing $M_{\rm UV}^{\rm lim}$ to much brighter values that are actually inconsistent with the steep slope of the observed UV luminosity function at high redshift. However, the IMF is appreciably more top heavy than the classic Salpeter or Chabrier shapes in that star masses $\lesssim 1\, M_\odot$ are appreciably suppressed. This may be related (but not exclusively) to the thermal Jeans limit from the hot CMB background.

\begin{figure}[H]
\includegraphics[width=0.51\textwidth]{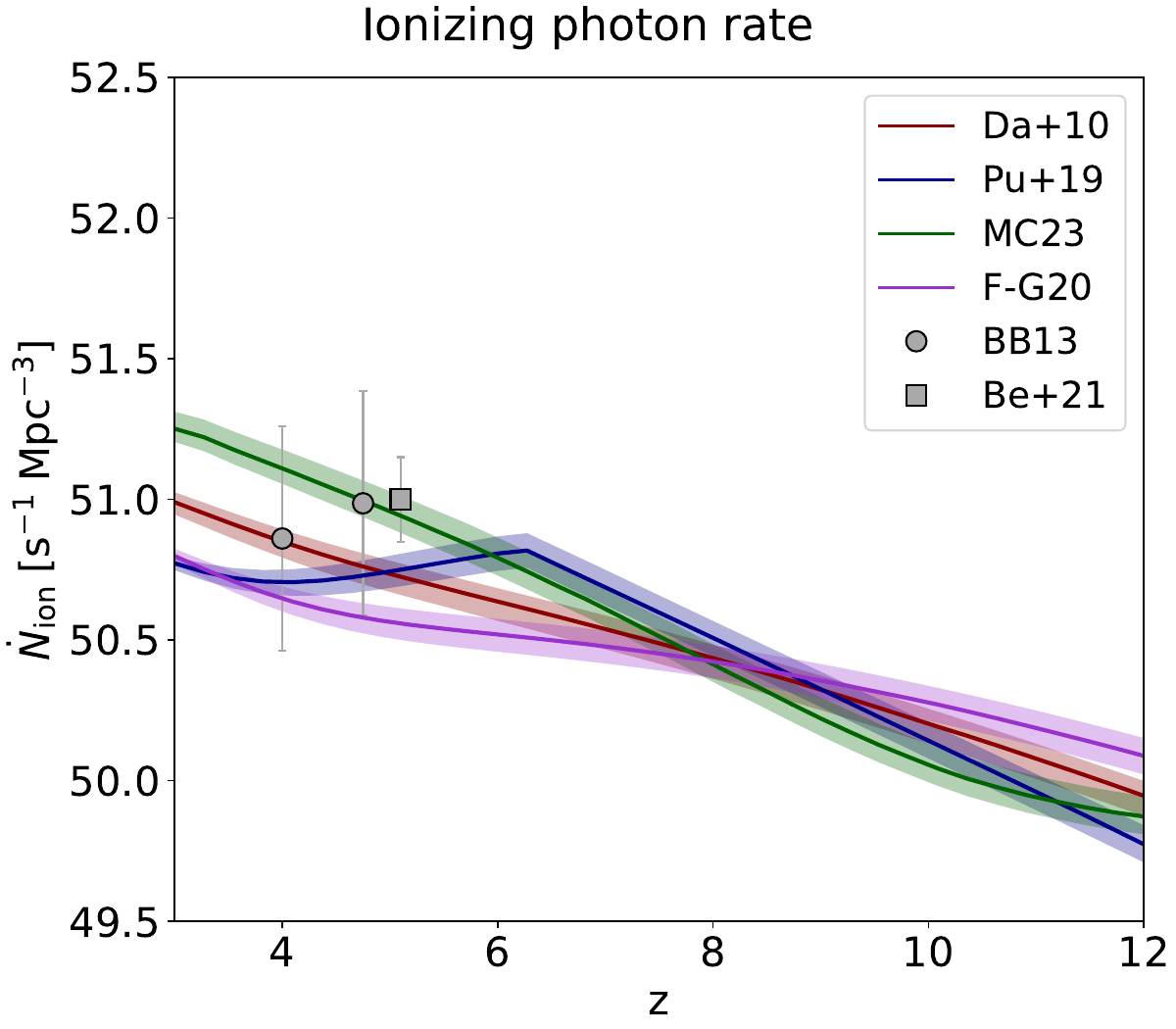}\\
\\
\includegraphics[width=0.51\textwidth]{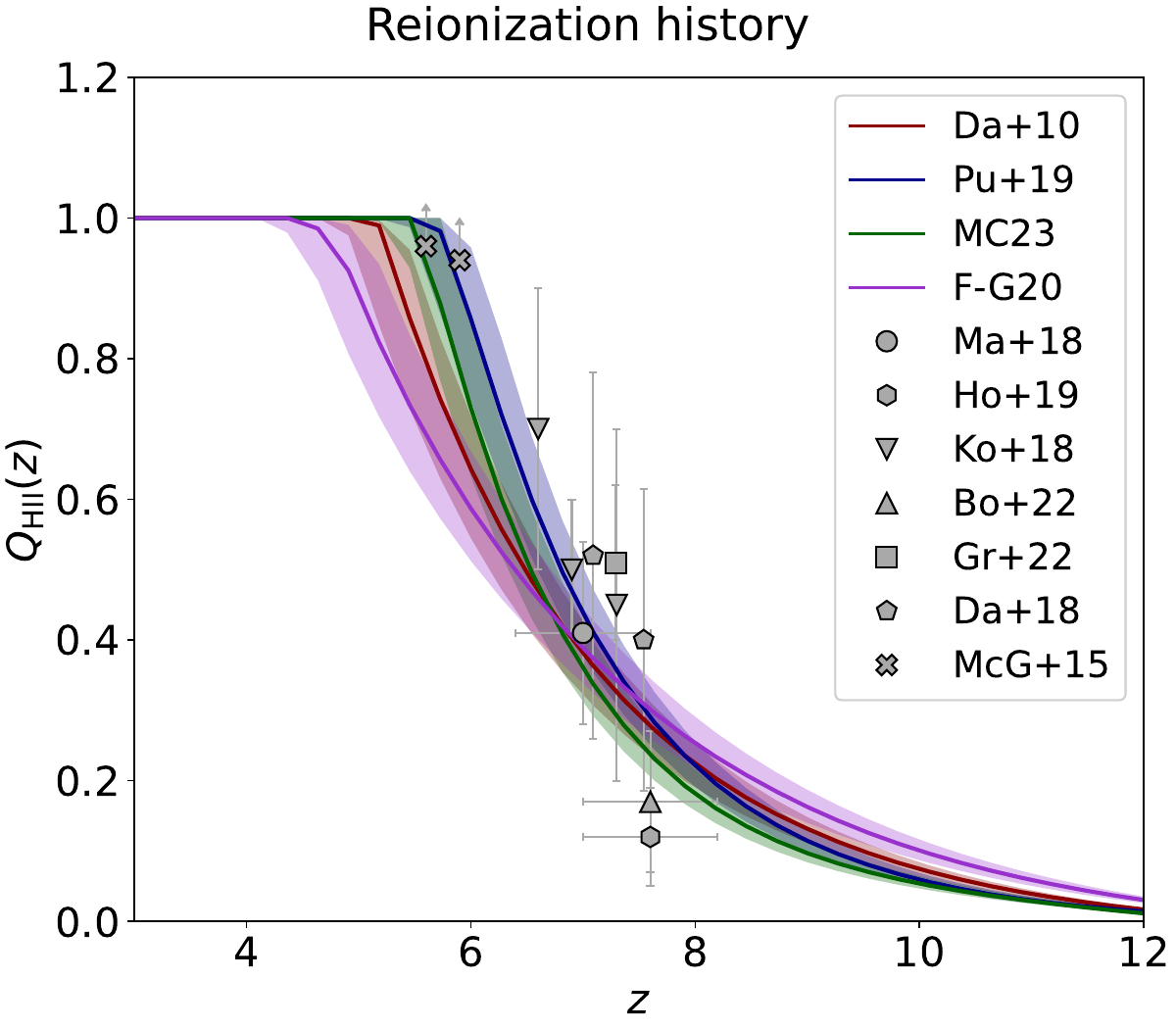}\\
\\
\includegraphics[width=0.51\textwidth]{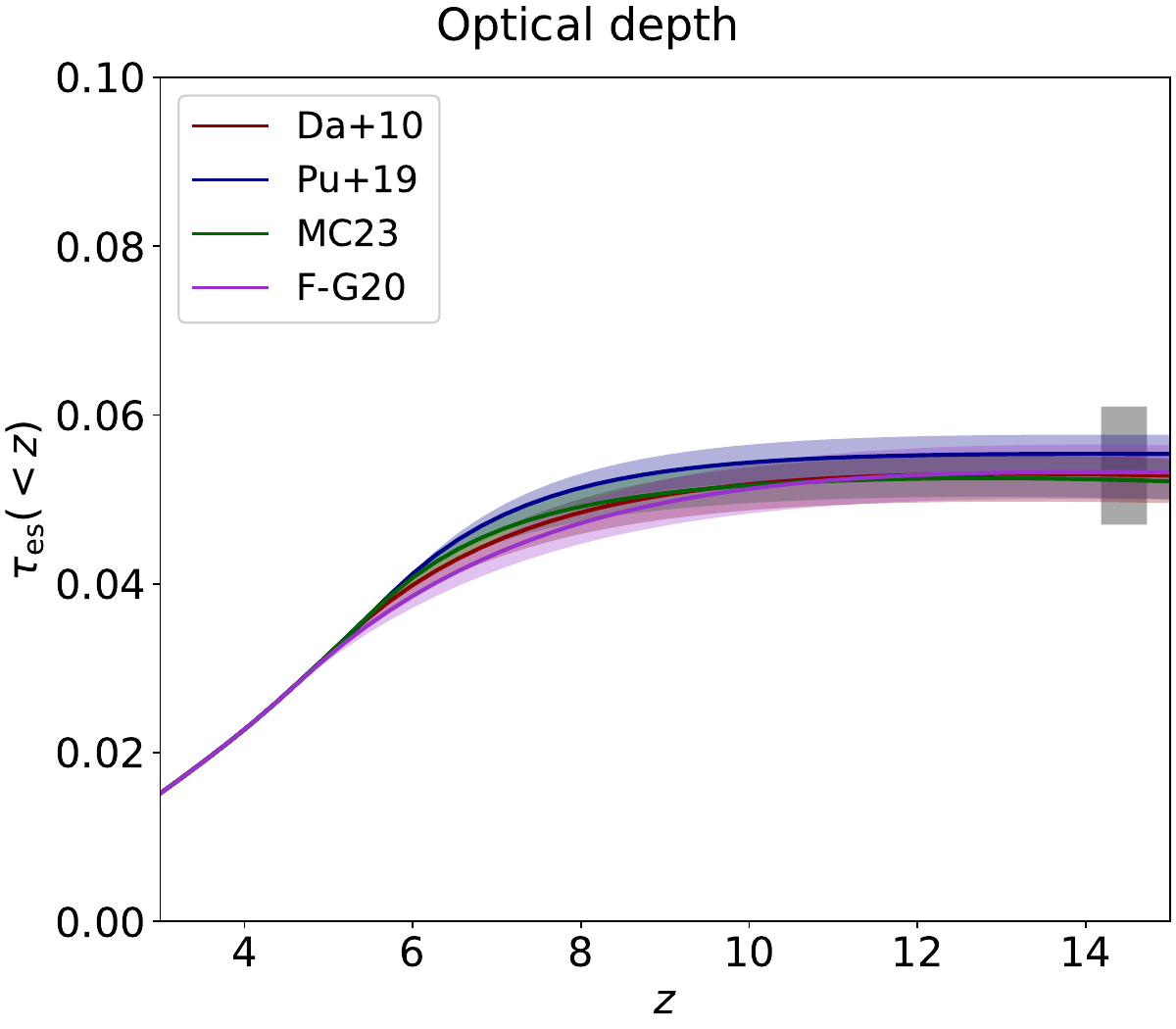}
\caption{Fits to the ionizing photon rate (top panel), reionization history in terms of the ionized hydrogen volume filling factor (middle panel), and optical depth for electron scattering (bottom panel). Colored lines illustrate the median, and the shaded areas show the $2\sigma$ credible interval from sampling the posterior distribution for each of the escape fraction parameterizations by \cite{Dayal20} (red), \cite{Puchwein19} (blue), \cite{Mitra23} (green), and \cite{Faucher20} (magenta). Data are from \cite{Becker13} (circles) and \cite{Becker21} (squares) in the top panel; \cite{Mason18} (circles), \cite{Hoag19} (hexagons), \cite{Konno14,Konno18} (reversed triangles), \cite{Bolan22} (triangles), \cite{Greig22} (squares), \cite{Davies18} (pentagons), and \cite{McGreer15} (crosses) in the middle panel; \cite{Aghanim20} (gray shaded area) in the bottom panel.}\label{fig|reion_best}
\end{figure}

In Figure \ref{fig|reion_best}, the best fit (solid lines) and the $2\sigma$ credible intervals (shaded areas) sampled from the posteriors are projected onto the reionization observables: the cosmic ionizing photon rate; the reionization history in terms of the redshift evolution of the volume filling factor for ionized hydrogen; and the electron scattering optical depth. The fits for all four escape fraction parameterizations are very good, as also testified by the reduced $\chi^2_r$ reported in Table \ref{tab|results}. There is a slight preference for escape fraction models by \cite{Mitra23,Puchwein19} over the ones \linebreak by \cite{Dayal20,Faucher20}; however, the differences in $\chi^2_r$ and in the likelihood evidence are not statistically significant enough to draw definite conclusions. It is fair to say that all escape fraction models perform comparably well in fitting reionization observables.

\begin{figure}[H]
\includegraphics[width=0.775\textwidth]{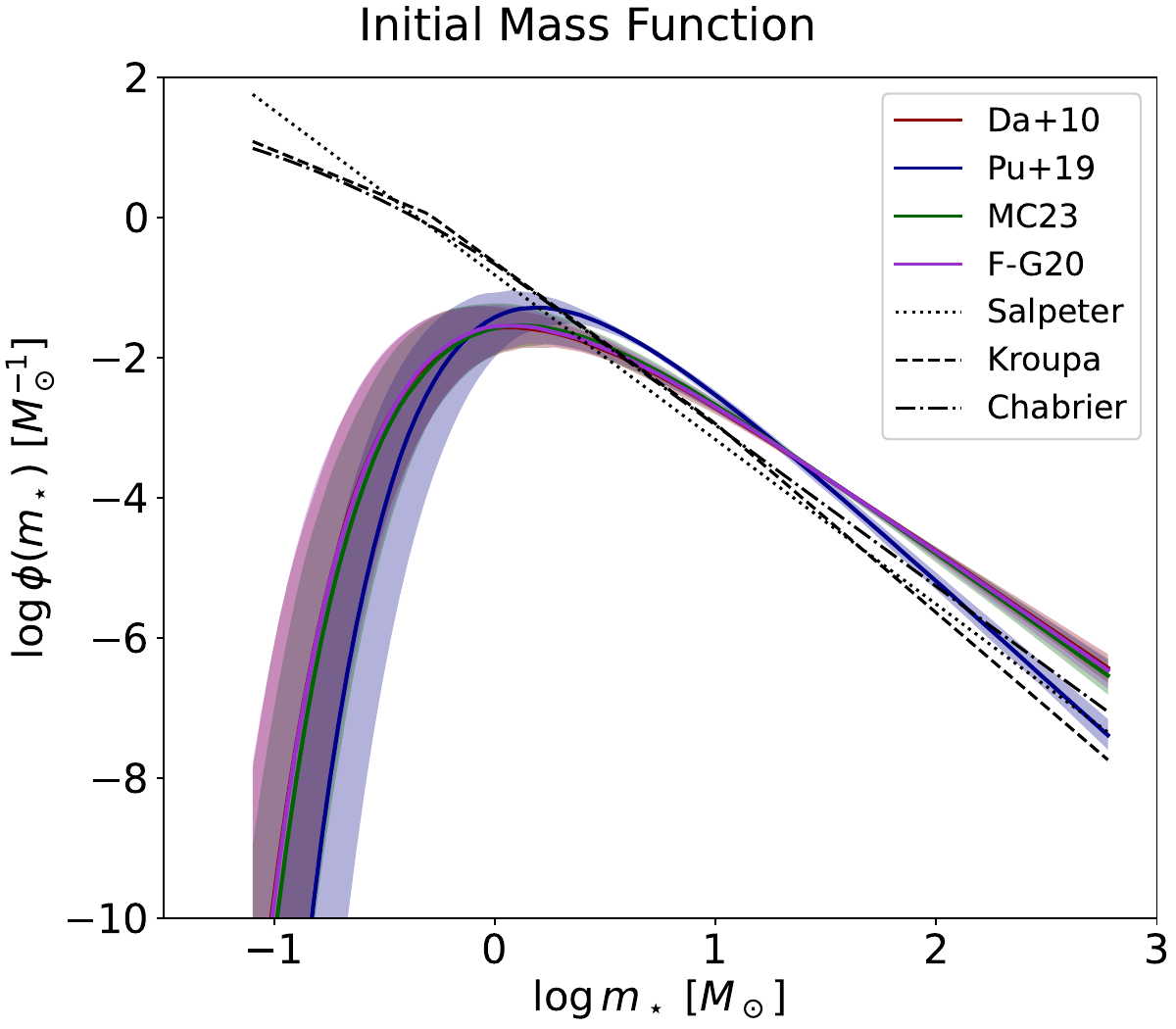}
\caption{IMF determination from our Bayesian analysis. Colored lines illustrate the median, and shaded areas show the $2\sigma$ credible interval from sampling the posterior distributions for each of the escape fraction parameterizations by \cite{Dayal20} (red), \cite{Puchwein19} (blue), \cite{Mitra23} (green), and \cite{Faucher20} (magenta). The Salpeter (dotted), Kroupa (dashed), and Chabrier (dot-dashed) IMFs are also plotted for reference.}\label{fig|IMF_best}
\end{figure}

Finally, in Figure \ref{fig|IMF_best}, we display the reconstructed IMF for the different escape fraction models. Within $2\sigma$ confidence intervals, all IMFs are very similar, with a slope $\xi\sim -2.2$ at high masses (slightly steeper $\xi\sim -2.8$ for the \cite{Puchwein19} escape fraction), a maximum at around a few $M_\odot$, and a cutoff at smaller mass, such that the number of stars with masses $\lesssim 1\, M_\odot$ is suppressed by a factor of $100$ or more with respect to a standard Salpeter or Chabrier shape. Comparing the overall shapes of the reconstructed IMF during the reionization epoch and the observed IMF in the local neighborhood leads to some physical conclusions. The IMF power law shape is the same and hence must originate by a fundamental process (e.g., turbulent fragmentation) in star-forming regions. The suppression in the formation of small stars, present at the reionization redshift but clearly at variance with the IMF in the local Universe, may be related to the vastly different environmental conditions (for example, the high thermal background provided by CMB photons in the reionization era).

%\begin{center}
%\end{center}

Clearly, such a reconstructed IMF with a paucity of small star masses below a few $M_\odot$ would have considerable consequences for the estimate of stellar masses from observed galaxy luminosities. This is particularly relevant in the light of the recent claims for substantial stellar masses estimated from JWST data at high redshift \cite{Labbe23,CurtisLake23,Finkelstein23}. In particular, puzzling cases are constituted by the galaxies from the \cite{Labbe23} sample, for which considerable stellar masses have been inferred from SED fitting to the JWST UV restframe photometry on adoption of a Salpeter IMF. Such galaxies have been shown to be marginally consistent with the standard $\Lambda$CDM cosmology \cite{BoylanKolchin23} in that, even assuming a maximal conversion of all the baryons into stars (star formation efficiency $\epsilon_\star\equiv M_\star/f_b\,M_{\rm H}\approx 1$), the stellar masses would correspond to extremely rare peaks in the density field.

%\begin{center}
%\end{center}

\begin{figure}[H]
\includegraphics[width=0.775\textwidth]{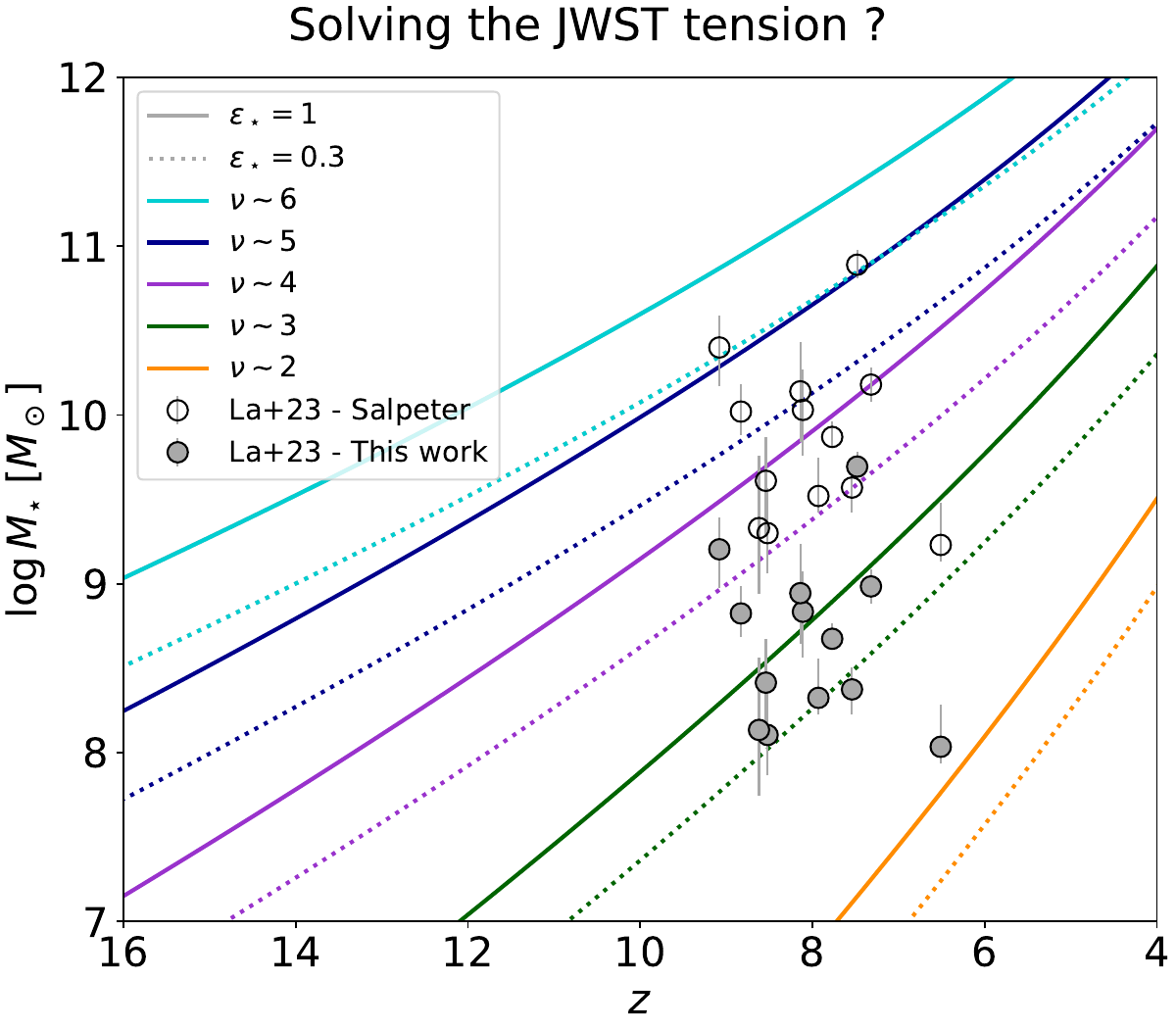}
\caption{Circles represent the estimate of stellar masses for the sample by \cite{Labbe23} inferred either with a Salpeter IMF (empty symbols) or with the IMF reconstructed in this work (filled symbols) under the \cite{Mitra23} escape fraction scenario. Colored lines illustrate the peak heights of the density field in terms of the ratio $\nu\equiv\delta_c/\sigma(M_{\rm H},z)=2$ (orange), $3$ (green), $4$ (magenta), $5$ (blue), and $6$ (cyan). Solid lines assume a maximal star formation efficiency $\epsilon_\star\approx 1$, while dotted ones refer to the typical value $\epsilon_\star\approx 0.3$ applying to local star-forming galaxies.}\label{fig|JWST}
\end{figure}

Basically, the related argument is illustrated in Figure \ref{fig|JWST}. Open symbols are the mass determinations from JWST data by \cite{Labbe23} with a Salpeter IMF. The colored lines represent different peak heights in the density field as quantified by the ratio $\nu\equiv \delta_c/\sigma(M_{\rm H},z)$ between the critical threshold for collapse $\delta_c \approx 1.7$ and the mass variance $\sigma(M_{\rm H},z)$ of the field smoothed (with a top-hat filter in real space) on the mass scale $M_{\rm H} =M_\star/(f_b\,\epsilon_\star)$ at redshift $z$ (typical peaks of the field should have $\nu\approx 1$). In addition, solid lines assume a maximal star formation efficiency $\epsilon_\star \approx 1$, while dotted ones refer to the typical value $\epsilon_\star \approx 0.3$ applying to local star-forming galaxies. It is seen that most of
the galaxies, and especially the two most massive ones, are extremely improbable in the standard $\Lambda$CDM cosmology, even assuming a maximal efficiency of star formation.

However, the situation improves substantially if we rely on the reconstructed IMF of this work. For the sake of simplicity, we exploit the outcomes for the escape fraction scenario by \cite{Mitra23}, but the results in the other cases are similar. At given UV luminosity and for the typical ages of the system at these high redshifts, the average $\langle M_\star/L_{\rm UV}\rangle$ ratios with our IMF are substantially reduced by a factor of $\gtrsim 10$ with respect to the Salpeter one as a consequence of the paucity of small stars below the characteristic mass. Therefore, at a given observed UV luminosity, the estimated galaxy stellar masses are reduced by a correspondingly large factor. Most of the galaxies are brought to values $\nu \lesssim 3.5$ and so are now in reasonable agreement with the $\Lambda$CDM expectations, even assuming a standard value of the star formation efficiency $\epsilon_\star\approx 0.3$. All in all, our reconstructed IMF appreciably alleviates the putative tension between JWST data and the standard cosmological framework.

%\begin{center}
%\end{center}

\section{Summary}\label{sec|summary}

In this paper, we have constrained the stellar initial mass function (IMF) during the epoch of reionization in the standard $\Lambda$CDM cosmology.

To this purpose, we have built up a semi-empirical model for the reionization history of the Universe based on various ingredients: the latest determination of the UV galaxy luminosity function from JWST out to redshift $z\lesssim 12$; {data-inferred and simulation-driven} assumptions on the redshift-dependent escape fraction of ionizing photons from primordial galaxies; a simple yet flexible parameterization of the IMF $\phi(m_\star)\sim m_\star^\xi\, e^{-m_{\star,\rm c}/m_\star}$ in terms of a slope $\xi<0$ at the high-mass end and of a characteristic mass $m_{\star,\rm c}$, below which it flattens or bends downwards; the \texttt{PARSEC} stellar evolution code to compute the UV and ionizing emission from different stars' masses as a function of age and metallicity; and a few physical constraints related to stellar and galaxy formation in faint galaxies at the \linebreak reionization redshifts.

Our model has three main parameters: the limiting magnitude $M_{\rm UV}^{\rm lim}$ down to which the luminosity function is steeply increasing, the IMF slope $\xi$, and characteristic mass $m_{\star,\rm c}$.  We have inferred these parameters via a standard MCMC technique, comparing our model outcomes with the reionization observables from different astrophysical and cosmological probes, including the cosmic ionizing photon rate, the redshift-dependent volume filling factor of ionized hydrogen, and the electron scattering optical depth from the \textit{Planck} mission.

We have found that the IMF slope $\xi$ is within the range from $-2.8$ to $-2.3$, consistent with direct determination from star counts in the Milky Way, while appreciably flatter slopes (hence, a strongly top-heavy IMF) are excluded at more than $5\sigma$ significance. The universality of the slope suggests that such a feature is originated by a fundamental process (e.g., turbulent fragmentation) in star-forming regions. On the other hand, the bestfit values of the IMF characteristic mass $m_{\star,\rm c}\sim$ a few $M_\odot$ implies a suppression in the formation of small stellar masses at variance with the local Universe. {This occurrence may be related, though not exclusively, to} the high thermal background of $\sim$ 20--30 K provided by CMB photons at the reionization redshifts. We have highlighted that such results are robust against different parameterizations for the redshift evolution of the escape fraction, which nevertheless remains the most uncertain ingredient in reionization studies. Observational campaigns and hydrodynamical simulations aimed at better reconstructing the escape fraction would be extremely welcome to strengthen the IMF constraints derived here.

Finally, we have investigated the implications of our reconstructed IMF for the recent JWST detections of massive galaxies at and beyond the reionization epoch. We have found that the stellar mass estimates from JWST data are considerably reduced when assuming the IMF from this work with respect to a Salpeter one. All in all, our reconstructed IMF substantially alleviates the tension of JWST data at high redshift with the standard cosmological framework.

\vspace{6pt}

\authorcontributions{Conceptualization: A.L., G.G., and L.B.; methodology: A.L., G.G., A.B., and B.S.H.; validation: F.G.,  M.M., B.S.H., C.B., A.B., and L.D.; writing: A.L., G.G., L.B., and F.G. All authors have read and agreed to the published version of the manuscript.}

\funding{This work was partially funded by the projects ``Data Science methods for MultiMessenger Astrophysics \& Multi-Survey Cosmology'', funded by the Italian Ministry of University and Research, Programmazione triennale 2021/2023 (DM no.2503, 9 December 2019), Programma Congiunto Scuole; EU H2020-MSCA-ITN-2019 no. 860744 {BiD4BESt: ``Big Data Applications for Black Hole Evolution Studies''}; Italian Research Center of High-Performance Computing Big Data and Quantum Computing (ICSC), funded by the European Union---NextGenerationEU---and National
Recovery and Resilience Plan (NRRP)---Mission 4 Component 2 within the activities of Spoke 3 (Astrophysics and Cosmos Observations); PRIN MUR 2022 project no. 20224JR28W ``Charting Unexplored Avenues in Dark Matter''; INAF Large Grant 2022 funding scheme with the project ``MeerKAT and LOFAR Team Up: a Unique Radio Window on Galaxy/AGN Co-Evolution''; INAF GO-GTO Normal 2023 funding scheme with the project ``Serendipitous H-ATLAS-fields Observations of Radio Extragalactic Sources (SHORES)''.}

\dataavailability{Data are contained within the article.}
%MDPI: This section is applied to all journal papers that contain original data; it is necessary for Article, Brief Report, Case Report, Communication, Data Descriptor, Letter, Proceeding Paper, and Project Report. Normally, the DAS for these research articles can state "Data are contained within the article." or "Data are contained within the article and supplementary materials." Please refer to the complete guideline at https://www.mdpi.com/ethics#_bookmark21. Please confirm whether the statement "Data are contained within the article." could be used here.
%AUTHORS: OK

\acknowledgments{{We acknowledge the three anonymous referees for their helpful comments and suggestions}. We warmly thank M. Talia and G. Rodighiero for illuminating discussions about high-redshift UV/NIR dark galaxies and JWST data.}

\conflictsofinterest{The authors declare no conflicts of interest.}

\begin{adjustwidth}{-\extralength}{0cm}

\reftitle{References}
%\printendnotes[custom]

\PublishersNote{}
\end{adjustwidth}

\end{document}